\documentclass[%
 reprint,
 superscriptaddress,
 preprintnumbers,
 nofootinbib,
 amsmath,amssymb,
 aps,
 onecolumn,
]{revtex4-2}

\usepackage{graphicx}% Include figure files
\usepackage{dcolumn}% Align table columns on decimal point
\usepackage{bm}% bold math
\usepackage{hyperref}% add hypertext capabilities
\usepackage{color}
\usepackage{braket}

\newcommand{\bA}{\mathbf{A}}
\newcommand{\be}{\mathbf{e}}

\newcommand{\bn}{\mathbf{n}}
\newcommand{\bx}{\mathbf{x}}
\newcommand{\pd}{\partial}
\newcommand{\obs}{\text{obs}}

\begin{document}

\preprint{KUNS-3069}

\title{Signature of polarized ultralight vector dark matter in pulsar timing arrays}% Force line breaks with \\
% \thanks{}%

\author{Kimihiro Nomura}
 \email{k.nomura@tap.scphys.kyoto-u.ac.jp}
 \affiliation{Department of Physics, Kyoto University, Kyoto 606-8502, Japan}%
 % \altaffiliation[Also at ]{}%Lines break automatically or can be forced with \\
\author{Hidetoshi Omiya}%
 \email{omiya@tap.scphys.kyoto-u.ac.jp}
 \affiliation{Department of Physics, Kyoto University, Kyoto 606-8502, Japan}%
\author{Takahiro Tanaka}%
 \email{t.tanaka@tap.scphys.kyoto-u.ac.jp}
 \affiliation{Department of Physics, Kyoto University, Kyoto 606-8502, Japan}%
 \affiliation{Center for Gravitational Physics and Quantum Information, Yukawa Institute for Theoretical Physics, Kyoto University, Kyoto 606-8502, Japan}

% \collaboration{}%\noaffiliation

% \author{}
%  \homepage{}
% \affiliation{}%
% \affiliation{}%
% \author{}
% \affiliation{}%

% \collaboration{}%\noaffiliation

\date{\today}% It is always \today, today,

\begin{abstract}
We investigate observational signatures of ultralight vector dark matter with masses $m \sim 10^{-24}$--$10^{-22}\,\text{eV}$ in pulsar timing arrays, taking into account general polarization states of the vector field.
We find that vector dark matter induces pulsar timing residuals with nontrivial directional dependence, reflecting the anisotropic property and polarization structure specific to vector dark matter, unlike scalar dark matter. 
We also derive angular correlation curves of the timing residuals. 
Intriguingly, circular polarization of the vector dark matter enhances the quadrupole nature of the correlation curve, resulting in a more notable bending of the Hellings--Downs curve.
The derived correlation curves offer a useful means to distinguish gravitational wave and dark matter contributions and to probe the nature of dark matter.
\end{abstract}

\maketitle

%\tableofcontents

\section{Introduction}
\label{sec:introduction}

The existence of an unknown non-baryonic matter component, referred to as dark matter (DM), is supported by many astronomical and cosmological observations. 
Among the compelling candidates for DM are ultralight bosonic fields, including scalar and vector fields, with masses below the electron volt scale \cite{Ferreira:2020fam, Hui:2021tkt, Nelson:2011sf, Arias:2012az}. 
In particular, fields with masses around $10^{-23}$--$10^{-22}\, \text{eV}$, also called fuzzy DM, are motivated to address small-scale challenges in the cold DM paradigm, such as the galactic core-cusp problem \cite{Hu:2000ke, Hui:2016ltb}.
A key feature of the ultralight DM scenario is that these bosons behave as classical fields due to their huge occupation numbers, and their coherent oscillations can effectively mimic cold DM on cosmological scales.

Pulsar Timing Arrays (PTAs) are experiments for detecting gravitational waves at nanohertz frequencies by measuring the arrival times of radio pulses from pulsars within our galaxy \cite{1990ApJ...361..300F}.
Recently, several PTA projects have reported evidence for a stochastic gravitational wave background \cite{EPTA:2023fyk, NANOGrav:2023gor, Reardon:2023gzh, Xu:2023wog}, stimulating momentum toward more robust detections.
Interestingly, PTAs can be used to search not only for gravitational waves but also for ultralight DM.
Oscillating coherently at a frequency set by its mass $m$, the ultralight DM field behaves like a fluid whose stress tensor oscillates at a frequency $f \approx 2m/(2\pi)$, thereby sourcing metric perturbations at the same frequency via Einstein’s equations.\footnote{
In addition to the ``fast'' mode with a frequency $f \approx 2m/(2\pi)$, a ``slow'' mode with $f \sim mv^2/(2\pi)$ can also be induced due to the velocity dispersion $v^2$ of DM, although the amplitude of the latter is suppressed \cite{Kim:2023kyy, Dror:2025nvg}. The present paper focuses only on the fast mode.
} 
The induced metric perturbations can, in turn, lead to potentially observable pulsar timing residuals.
Given that PTAs are sensitive to frequencies in the nanohertz band, they can probe ultralight DM with masses $m \sim 10^{-24}$--$10^{-22}\, \text{eV}$.
This idea was first proposed by Khmelnitsky and Rubakov \cite{Khmelnitsky:2013lxt}, in which the pulsar timing residuals induced by scalar field DM are calculated. 
Subsequent studies extended this idea to vector \cite{Nomura:2019cvc} and tensor field DM \cite{Armaleo:2020yml, Wu:2023dnp}, demonstrating that the resulting timing residuals exhibit anisotropies due to the directional nature of these fields, unlike the isotropic signature expected from a scalar field.
Based on these predictions, constraints on the DM abundance have been obtained using PTA observations \cite{Porayko:2014rfa, Porayko:2018sfa, Kato:2019bqz, PPTA:2022eul, Xia:2023hov, NANOGrav:2023hvm, EuropeanPulsarTimingArray:2023egv}.
Furthermore, the angular correlation curve of the timing residuals between different pulsars induced by vector DM, which is an analog of the Hellings--Downs curve in the case of stochastic gravitational waves \cite{Hellings:1983fr}, was calculated in Ref.~\cite{Omiya:2023bio} assuming that the vector field is linearly polarized, i.e., oscillating in a single direction.
Due to this directional property, the resulting correlation exhibits a characteristic angular dependence.
The angular correlation curve is also calculated for tensor DM \cite{Cai:2024thd}.\footnote{References~\cite{Armaleo:2020yml, Cai:2024thd} focus on the effects of direct coupling between tensor DM and ordinary matter, whereas Ref.~\cite{Wu:2023dnp} investigates the purely gravitational effects of tensor DM, which is the counterpart to our analysis for vector DM.}
Determining the angular correlation is crucial for distinguishing the contribution of ultralight DM from that of gravitational waves \cite{Arjona:2024cex}.

A vector DM has a more distinctive feature different from a scalar DM: the vector field can be polarized \cite{Zhang:2021xxa, Jain:2021pnk} (see also Refs.~\cite{Amin:2022pzv, Amaral:2024tjg}).
If all components of the vector field oscillate with the same phase, the field is linearly polarized; i.e., it oscillates along a single spatial direction. In contrast, if different components oscillate with different phases, the field exhibits elliptical polarization; i.e., the tip of the vector draws an ellipse.
Even if the vector DM were initially produced in a linearly polarized state in the early universe, it is natural to expect that subsequent nonlinear evolution, including halo mergers, would generally lead to elliptical polarization.
Therefore, to comprehensively search for the vector DM through observations, it is essential to predict the signatures associated with general polarization states.

In this paper, we investigate observational signatures of ultralight vector DM in PTAs, taking into account general polarization states of the vector field. This extends previous analyses \cite{Nomura:2019cvc, Omiya:2023bio} that considered only the linearly polarized case.
Given that the coherence time of the ultralight DM field exceeds $10^6\,\text{yr}$ for masses of interest, $m \lesssim 10^{-22}\,\text{eV}$, a specific polarization state is expected to persist over observational timescales.
The polarization state of the vector DM can then be characterized by its ellipticity.
We compute the pulsar timing residuals and their angular correlations induced by the vector DM, and clarify their dependence on the polarization state.\footnote{Reference \cite{Dror:2025nvg} also investigates the PTA signal from vector DM, but differs from our analysis in that an ensemble average over various polarization states is considered there.}
This study provides a more comprehensive set of signal predictions for probing the vector DM in PTA experiments.

This paper is organized as follows.
In Sec.\,\ref{sec:vectorDM}, we introduce ultralight vector DM and describe its polarization states.
In Sec.\,\ref{sec:metric_perturbation}, we compute metric perturbations induced by the vector DM.
In Sec.~\ref{sec:pulsar_timing}, we show that the metric perturbations modify the arrival times of pulses from pulsars.
In Sec.~\ref{sec:angular_cor}, we compute the angular correlation of the pulsar timing residuals induced by the vector DM, and reveal that the correlation pattern varies depending on the polarization state.
Section \ref{sec:conclusion} is devoted to the conclusion.
We adopt the natural units, in which the speed of light $c$ and the reduced Planck constant $\hbar$ are set to unity.

\section{Ultralight vector DM}
\label{sec:vectorDM}

\subsection{The model}

In this paper, we consider a vector DM field denoted by $A_\mu$. The action is given by 
\begin{align}
    S[A_\mu] = \int d^4 x \, \sqrt{-g} \,
    \bigg( - \frac{1}{4} g^{\mu\rho} g^{\nu\sigma} F_{\mu\nu} F_{\rho\sigma} - \frac{1}{2} m^2 g^{\mu\nu} A_\mu A_\nu \bigg) \, , 
    \label{eqaction1}
\end{align}
where $g$ represents the determinant of the metric $g_{\mu\nu}$, $F_{\mu\nu} \equiv \pd_\mu A_\nu -\pd_\nu A_\mu$ is the 
field strength tensor of the vector field, and $m$ is the mass of the vector field. 
In this section, we first describe the DM configuration in the galaxy.
The discussion here corresponds to a natural extension of the scalar DM case \cite{Hui:2016ltb, Ferreira:2020fam, Hui:2021tkt} to the vector case \cite{Nomura:2019cvc}.

As we consider the galactic scale, the cosmic expansion is neglected.
Additionally, assuming that the gravitational potential is shallow, the background metric can be approximated as flat.
Taking the variation of the action \eqref{eqaction1}, the equation of motion for the vector field $A_\mu$ in the flat space reads 
\begin{align}
    \pd_\nu F^{\nu\mu} - m^2 A^\mu = 0 \,. 
    \label{eqvdmbg4}
\end{align}
Using the identity $\pd_\mu \pd_\nu F^{\nu \mu} = 0$, Eq.~\eqref{eqvdmbg4} is equivalent to the following set of equations,
\begin{align}
    \pd_\mu A^\mu &= 0 \,,
    \label{eqvdmbg5}\\
    (\pd_\nu \pd^\nu - m^2) A_\mu &= 0 \,.
    \label{eqvdmbg6}
\end{align}
Equation \eqref{eqvdmbg6} indicates that each component of $A_\mu$ obeys the Klein--Gordon equation with mass $m$. 
Considering a non-relativistic field with a small velocity $v \ll 1$, which is appropriate for ultralight DM fields, 
the angular frequency $\omega$ and wave number $k$ are approximated as $\omega \approx m$ and $k \approx mv \ll m$, respectively.
Then, Eq.~\eqref{eqvdmbg5} leads to $|A_0|/ |\bA| \approx k/\omega \ll 1$, where $\bA = (A_j)$ $(j=1,2,3)$ denotes the three-dimensional spatial vector.
Hence, the temporal component $A_0$ is suppressed compared to the spatial components $A_j$. 
The solution for the spatial components is represented as the oscillating field with angular frequency equal to the field mass $m$,
\begin{align}
    A_j (t, \bx) = \bar{A}_j(\bx) \cos (mt - \alpha_j(\bx)) \,. 
    \qquad 
    (\text{No sum over} ~ j.)
    \label{eqvdmbg1}
\end{align}
Here, $\bar{A}_j$ and $\alpha_j$ are real quantities corresponding to the amplitudes and phases of the oscillation, respectively.
In general, the amplitudes and phases may vary in both time and space, but their variation occurs on scales much longer than the oscillation timescale $m^{-1}$.
Indeed, they are almost constant over time and length scales shorter than the coherence time $\tau_{\text{coh}}$ and coherence length $\lambda_{\text{coh}}$, which are roughly estimated as 
\begin{align}
    \tau_{\text{coh}} &\sim \frac{2\pi}{mv^2} \sim 1 \times 10^7 \, \text{yr} \, \bigg( \frac{10^{-23}\, \text{eV}}{m} \bigg) \bigg( \frac{10^{-3}}{v} \bigg)^2 \,, 
    \label{tau_coh}\\
    \lambda_{\text{coh}} &\sim \frac{2\pi}{mv} \sim 4 \times 10^3 \, \text{pc} \, \bigg( \frac{10^{-23}\, \text{eV}}{m} \bigg) \bigg( \frac{10^{-3}}{v} \bigg) \,,
    \label{lambda_coh}
\end{align}
where $v \sim 10^{-3}$ is used as a typical velocity.
Since $\tau_{\text{coh}}$ is much longer than the observational timescale in PTAs, $T_{\text{obs}} \sim 10^{1}$--$10^{2}\,\text{yr}$, for DM masses of interest, both the amplitudes and phases can be regarded as time-independent. 
On the other hand, given that pulsars are distributed over a region of order kiloparsec \cite{EPTA:2023sfo, NANOGrav:2023pdq}, it is subtle whether the spatial variations can be neglected on the PTA length scale.
Therefore, we keep the spatial dependence of the amplitudes and phases in Eq.~\eqref{eqvdmbg1}.
For later convenience, we introduce the complex amplitude $\hat{A}_j \equiv \bar{A}_j e^{i \alpha_j}$ (no sum over $j$) for each direction to rewrite Eq.~\eqref{eqvdmbg1} as 
\begin{align}
    A_j(t, \bx) &= \text{Re} \big[ \hat{A}_j(\bx) \, e^{-imt} \big] \,,
    \label{eqvdmbg1-2}
\end{align}
where $\text{Re}[\cdots]$ stands for the real part.

From the action \eqref{eqaction1}, the energy-momentum tensor of the vector field is given by 
\begin{align}
    T_{\mu\nu} &\equiv \frac{-2}{\sqrt{-g}} \frac{\delta S}{\delta g^{\mu\nu}} 
    \notag \\
    &= g^{\rho\sigma} F_{\mu\rho} F_{\nu\sigma} - \frac{1}{4} g_{\mu\nu} g^{\rho\alpha} g^{\sigma \beta}F_{\alpha\beta} F_{\rho\sigma} 
    + m^2 A_\mu A_\nu
    - \frac{1}{2}  m^2 g_{\mu\nu} g^{\rho\sigma} A_\rho A_\sigma \,.
    \label{eqEMT1}
\end{align}
Substituting the configuration \eqref{eqvdmbg1} and neglecting the spatial gradient of the field, the leading components of the energy-momentum tensor are given by
\begin{align}
    T_{00} &= \frac{1}{2} m^2 \hat{A}^*_j \hat{A}^j 
    = \frac{1}{2} m^2 \bar{A}_j \bar{A}^j 
    \equiv \rho_{A}\,, 
    \label{eqT00-2}\\
    T_{0j} = T_{j0} &= 0 \,, 
    \label{eqT0j-2}\\
    T_{jk} &= m^2 \, \text{Re} \big[ \hat{A}_j \hat{A}_k e^{-2imt} \big] 
    - \frac{1}{2} m^2 \delta_{jk} \, \text{Re} \big[ \hat{A}_l \hat{A}^l e^{-2imt} \big] \,. 
    \label{eqTjk-2}
\end{align}
Here, the asterisk (*) represents the complex conjugate.
Note that the $T_{00} (\equiv \rho_{A})$ component gives the energy density of the vector field, which is almost constant in time. 
On the other hand, the $T_{jk}$ component includes a factor of $e^{-2imt}$, indicating that the pressure of the field oscillates with angular frequency of $2m$.

\subsection{Polarized states}
\label{subsec:polarized_state}

As described in Eq.~\eqref{eqvdmbg1}, the vector DM configuration consists of oscillations along the three spatial directions.
In general, the components in different spatial directions may have different phases.
Depending on their relative phases and amplitudes, various polarization states of the vector DM can be realized.

To describe the polarization of the vector DM, it is convenient to redefine the coordinate axes as follows. 
First, because of approximate angular momentum conservation, the vector $\bA(t,\bx) = (A_j(t,\bx))$ at a given point $\bx$ can be assumed to remain confined to a two-dimensional plane
(see Appendix \ref{app:Noether} for the definition of angular momentum).
We refer to this two-dimensional plane as the polarization plane.
Taking the $z$-axis to align with the direction of the angular momentum vector and identifying the $x$-$y$ plane with the polarization plane, the vector $\bA(t,\bx)$ at the point $\bx$ is expressed as 
\begin{align}
    \bA(t,\bx) = \bar{A}_x(\bx) \cos (mt - \alpha_x(\bx)) \, \mathbf{e}^x(\bx) + \bar{A}_y(\bx) \cos (mt - \alpha_y(\bx))
    \, \mathbf{e}^y(\bx) \,,
    \label{pol1}
\end{align}
where $\bar{A}_x$ and $\bar{A}_y$ are real amplitudes, $\alpha_x$ and $\alpha_y$ are real phases, and $\mathbf{e}^x$ and $\mathbf{e}^y$ are the local basis vectors along the $x$- and $y$-axes, respectively.
(The vectors $\mathbf{e}^x$ and $\mathbf{e}^y$ can be taken arbitrarily in the polarization plane as long as they are orthonormal.)
Note that the $z$-axis is defined by the direction of the angular momentum, which may vary in space on the scale of $\lambda_{\text{coh}}$, and accordingly the basis vectors $\mathbf{e}^x$ and $\mathbf{e}^y$ can also depend on the position $\bx$.

\begin{figure}[tb]
\centering
\includegraphics[width=90mm]{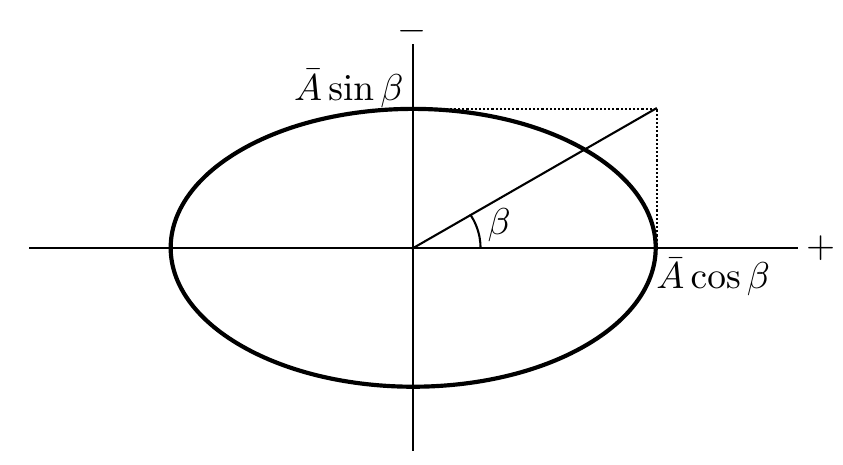}
\caption{The ellipse traced by the vector $\bA(t,\bx)$ in Eq.~\eqref{pol2}. The $+$ and $-$ axes are aligned with the principal axes of the ellipse.}
\label{fig:ellipse1}
\end{figure}

The tip of the vector \eqref{pol1} traces an ellipse in the polarization plane as time evolves. 
Indeed, as shown in Appendix \ref{app:ellipse}, by rotating the coordinate axes within the polarization plane, the vector can be represented in the following form,\footnote{
One can shift the origin of the time coordinate $t=0$ to eliminate the phase $\alpha(\bx)$ in Eq.~\eqref{pol2} at a given point $\bx$. We retain the phase $\alpha(\bx)$ since, in what follows, we consider the vector field values at different locations (Earth and pulsars), where the phases may differ.
} 
\begin{align}
    \bA(t,\bx) = \bar{A}(\bx) \cos \beta(\bx) \cos (mt - \alpha(\bx)) \, \mathbf{e}^+(\bx) + \bar{A}(\bx) \sin \beta(\bx) \sin (mt - \alpha(\bx)) \, \mathbf{e}^-(\bx) \,,
    \label{pol2}
\end{align}
where $\bar{A}$, $\beta$, and $\alpha$ are real quantities, and $\mathbf{e}^+$ and $\mathbf{e}^-$ are orthonormal basis vectors aligned with the principal axes of the ellipse.
The relations between $(\bar{A}, \beta, \alpha)$ in Eq.~\eqref{pol2} and $(\bar{A}_x, \bar{A}_y, \alpha_x, \alpha_y)$ in Eq.~\eqref{pol1} are given in Appendix \ref{app:ellipse}.
The ellipse traced by the tip of the vector \eqref{pol2} is illustrated in Fig.~\ref{fig:ellipse1}.
The vector rotates around the ellipse with a period $T = 2\pi/m$.
The angle $\beta$ (defined within $-\pi/2\leq \beta \leq \pi/2$) characterizes the ellipticity of the polarization, since the ratio of the axis lengths is given by $|\tan \beta|$.
For $\beta = 0$ or $\pm \pi/2$, the vector oscillates along a single direction, $\mathbf{e}^+$ or $\mathbf{e}^-$; thus, the vector exhibits linear polarization. 
For $\beta = \pm \pi/4$, the lengths of the principal axes become equal, and, thus, the vector exhibits perfect circular polarization.\footnote{
    Note that a circularly polarized state of the vector field can be realized by superposing linearly polarized waves.
    Therefore, a circularly polarized state of the vector DM does not necessarily need to be produced via a mechanism involving parity violation in the early universe.
    For example, vector DM produced in linearly polarized states with finite correlation lengths could subsequently evolve into a circularly polarized state through processes such as halo mergers.
}
When $\sin \beta$ is positive, the vector rotates counterclockwise around the ellipse; when $\sin \beta$ is negative, it rotates clockwise.

Equation \eqref{pol2} is rewritten in the form of Eq.~\eqref{eqvdmbg1-2} as $\bA(t, \bx) = \text{Re} [ \hat{\bA}(\bx) \, e^{-imt} ]$, where the complex vector $\hat{\bA}(\bx)$ is given by 
\begin{align}
    \hat{\bA} (\bx) &= \bar{A} (\bx) \, e^{i\alpha(\bx)} \big[ \mathbf{e}^+ (\bx) \cos \beta (\bx) + i \,\mathbf{e}^- (\bx) \sin \beta (\bx) \big]\,. 
    \label{pol4}
\end{align}

\section{Metric perturbation induced by vector DM}
\label{sec:metric_perturbation}

In this section, we find metric perturbations induced by the oscillating vector DM.
The induced oscillating metric perturbation affects the arrival time of pulses from pulsars, as seen in the next section.
We write the spacetime metric $g_{\mu\nu}$ as the sum of the Minkowski metric $\eta_{\mu\nu}$ and a small perturbation $h_{\mu\nu}$:
\begin{align}
    g_{\mu\nu} = \eta_{\mu\nu} + h_{\mu\nu}\,. 
    \label{metric1}
\end{align}
We adopt the harmonic gauge for the metric perturbation, 
\begin{align}
    \pd^\nu \bar{h}_{\mu\nu} = 0\,, 
    \label{metric2}
\end{align}
where $\bar{h}_{\mu\nu} \equiv h_{\mu\nu} - (1/2) \eta_{\mu\nu} h$ with $h \equiv \eta^{\mu\nu} h_{\mu\nu}$. 
In this gauge, the Einstein equation is represented as 
\begin{align}
    \partial^\lambda \partial_\lambda \bar{h}_{\mu\nu} &= - 16 \pi G \, T_{\mu\nu}\,, 
    \label{metric3}
\end{align}
where $G$ is the Newton constant.
We can substitute the energy-momentum tensor \eqref{eqT00-2}--\eqref{eqTjk-2} into the right-hand side to find the metric perturbation induced by the vector DM.
For the spatial components $(\mu\nu = jk)$, the energy-momentum tensor $T_{jk}$ \eqref{eqTjk-2} is oscillating in time and its spatial dependence is suppressed. 
Therefore, the Einstein equation is approximated as $\pd_0^2 \bar{h}_{jk} = 16 \pi G \, T_{jk}$, and it can be integrated as 
\begin{align}
    \bar{h}_{jk}(t, \bx) 
    &= - 4 \pi G \, \text{Re} \big[ \hat{A}_j(\bx) \hat{A}_k(\bx) \, e^{-2imt} \big] 
    + 2 \pi G \, \delta_{jk} \, \text{Re} \big[ \hat{A}_l(\bx) \hat{A}^l(\bx) \, e^{-2imt} \big]\,. 
    \label{metric4}
\end{align}
The gauge condition \eqref{metric2} leads to $\pd^0 \bar{h}_{j0} = - \pd^k \bar{h}_{jk}$ and $\pd^0 \bar{h}_{00} = - \pd^j \bar{h}_{0j}$. 
Due to suppression of the spatial gradient, these imply that the components $\bar{h}_{00}$ and $\bar{h}_{0j}$ can be neglected compared to $\bar{h}_{jk}$ when considering the time-oscillating parts relevant for PTA observations.
Then, the spatial metric perturbation $h_{ij}$ can be found through $h_{jk} = \bar{h}_{jk} - (1/2) \delta_{jk} \bar{h}$ with $\bar{h} = \eta^{\mu\nu} \bar{h}_{\mu\nu}$ as 
\begin{align}
    h_{jk}(t,\bx) = - 4 \pi G \,\text{Re} \big[ \hat{A}_j(\bx) \hat{A}_k(\bx) \, e^{-2imt} \big] 
    + \pi G \, \delta_{jk} \, \text{Re} \big[ \hat{A}_l(\bx) \hat{A}^l(\bx) \, e^{-2imt} \big]\,,
    \label{metric5}
\end{align}
where only the oscillating parts are shown.
In principle, the other components $h_{00}$ and $h_{0j}$ can also be computed, but the spatial components $h_{ij}$ are sufficient to evaluate the effects on pulsar timing \cite{Khmelnitsky:2013lxt, Nomura:2019cvc, Dror:2025nvg}.

\section{Effects on pulsar timing}
\label{sec:pulsar_timing}

The oscillating metric perturbations induced by the vector DM affect the arrival time of pulses emitted from pulsars.
We define the redshift of the pulses from the pulsar $a$ as 
\begin{align}
    z_a(t) \equiv \frac{\omega_{0,a} - \omega_{\obs,a}(t)}{\omega_{0,a}} \,,
    \label{eqz1}
\end{align}
where $\omega_{0,a}$ is the intrinsic angular frequency of the pulsar $a$, and $\omega_{\obs, a}(t)$ is the angular frequency of the arrival of the pulses from the pulsar $a$ observed on the Earth at time $t$.
In other words, $\omega_{\obs,a}(t) \equiv 2\pi/ T_{\obs,a}(t)$, where $T_{\obs,a}(t)$ represents the arrival period of the pulses from the pulsar $a$ at time $t$.

Given the metric perturbation \eqref{metric5}, the leading contribution to the redshift is calculated as \cite{Detweiler:1979wn, Book:2010pf, Nomura:2019cvc, Dror:2025nvg}
\begin{align}
    z_a(t) 
    &= \frac{1}{2} n_a^j n_a^k \, [ h_{jk}(t, \bm{0}) - h_{jk}(t-L_a, \bx_a) ] \,.
    \label{eqz2}
\end{align}
Here, the origin of the spatial coordinate $\bx = \bm{0}$ is chosen at the observation point (Earth).
The position of the pulsar $a$ is denoted by $\bx_a = L_a \bn_a$, where $L_a$ is the distance to the pulsar $a$ and $\bn_a = (n_a^j)$ is the unit vector pointing to the pulsar $a$.
From the expression for the metric perturbation induced by the vector DM \eqref{metric5}, the redshift is represented as 
\begin{align}
    z_a(t) 
    &= z_{a,\text{E}} (t) - z_{a,\text{P}} (t) \,,
    \label{eqz4}
\end{align}
where the first term represents the ``Earth term'' given by
\begin{align}
    z_{a, \text{E}}(t) &= 
    - 2\pi G \, n_a^j n_a^k \, \text{Re} \big[ \hat{A}_j(\bm{0}) \hat{A}_k(\bm{0}) \, e^{-2imt} \big] 
    + \frac{\pi G}{2} \, \text{Re} \big[ \hat{A}_j(\bm{0})  \hat{A}^j(\bm{0}) \, e^{-2imt} \big] \,,
    \label{eqz5}
\end{align}
and the latter represents the ``pulsar term'' given by
\begin{align}
    z_{a, \text{P}}(t) &= 
    - 2\pi G \, n_a^j n_a^k \, \text{Re} \big[ \hat{A}_j(\bx_a) \hat{A}_k(\bx_a) \, e^{-2im(t-L_a)} \big]
    + \frac{\pi G}{2} \, \text{Re} \big[ \hat{A}_j(\bx_a)  \hat{A}^j(\bx_a) \, e^{-2im(t-L_a)} \big]
     \,.
    \label{eqz6}
\end{align}
Note that the field values of the vector DM at the Earth and pulsar positions determine the Earth and pulsar terms, respectively.

\begin{figure}[tb]
\centering
\includegraphics[width=\textwidth]{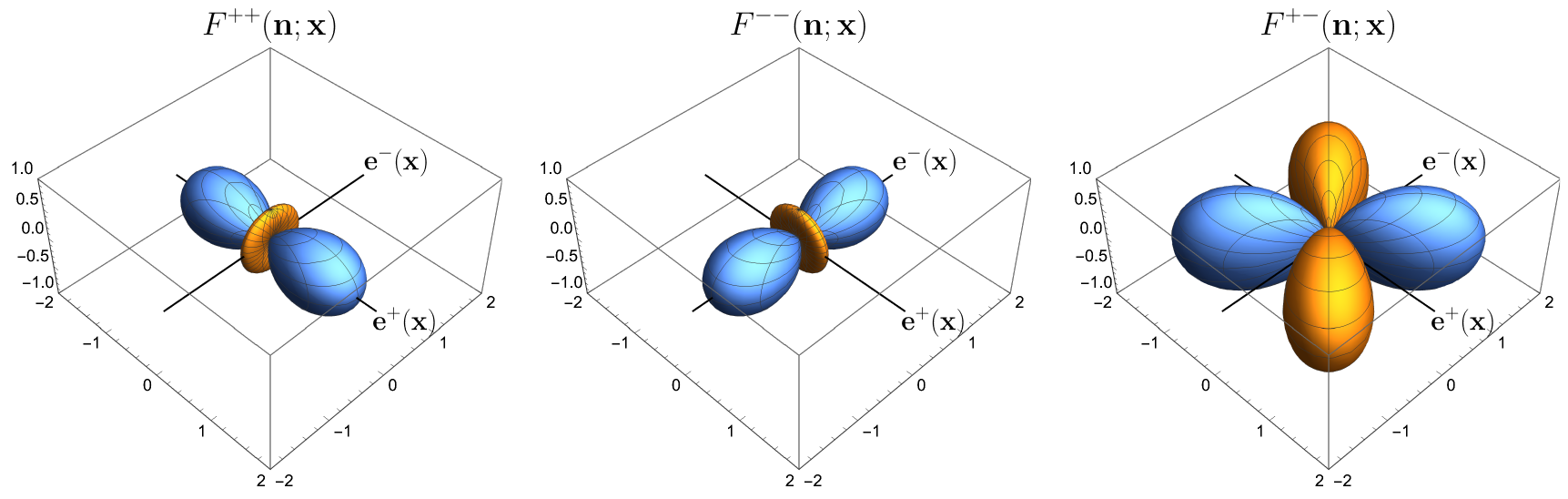}
\caption{The dependence of the pattern functions $F^{++}(\bn; \bx)$ (left), $F^{--}(\bn; \bx)$ (center), and $F^{+-}(\bn; \bx)$ (right) on the direction $\mathbf{n}$ is illustrated. The yellow and blue regions indicate where the function takes positive and negative values, respectively.}
\label{fig:pattern}
\end{figure}

As described in Sec.~\ref{subsec:polarized_state}, at a given point $\bx$, the vector ${A}_j(t, \bx)$ traces an ellipse in a two-dimensional polarization plane.
Following Eq.~\eqref{pol2}, we introduce two orthonormal basis vectors $\mathbf{e}^+ = (e^+_j)$ and $\mathbf{e}^- = (e^-_j)$ on the polarization plane at a given point, which are aligned with the principal axes of the ellipse.
Using these basis vectors, we express the complex vector $\hat{A}_j(\bx)$ at the point $\bx$ in the form of Eq.~\eqref{pol4}, i.e.,
\begin{align}
    \hat{A}_j(\bx) &= \hat{A}_+(\bx) \, e^+_j(\bx) + \hat{A}_-(\bx) \, e^-_j(\bx) \,,
    \label{eqdec1}
\end{align}
with
\begin{align}
    \hat{A}_+(\bx) &= \bar{A}(\bx) \, e^{i \alpha (\bx)} \cos \beta (\bx) \,, 
    \label{eqdec2}\\
    \hat{A}_-(\bx) &= \bar{A}(\bx) \, e^{i \alpha (\bx)} \,i \sin \beta (\bx) \,.
    \label{eqdec3}
\end{align}
Substituting Eq.~\eqref{eqdec1} into Eqs.~\eqref{eqz5} and \eqref{eqz6}, the Earth and pulsar terms of the redshift are expressed as 
\begin{align}
    z_{a,\text{E}}(t) &= \pi G \, \Big\{
        F^{++}(\bn_a; \mathbf{0}) \, \text{Re} \big[ \hat{A}_+^2 (\mathbf{0}) \, e^{-2imt} \big] 
        + F^{--}(\bn_a; \mathbf{0}) \, \text{Re} \big[ \hat{A}_-^2 (\mathbf{0}) \, e^{-2imt} \big] 
        \notag \\
        &\qquad\quad + F^{+-}(\bn_a; \mathbf{0}) \, \text{Re} \big[ \hat{A}_+ (\mathbf{0}) \, \hat{A}_- (\mathbf{0}) \, e^{-2imt} \big] 
    \Big\} \,,
    \label{eqzE1}\\
    z_{a,\text{P}}(t) &= \pi G \, \Big\{
        F^{++}(\bn_a; \bx_a) \, \text{Re} \big[ \hat{A}_+^2 (\bx_a) \, e^{-2im (t - L_a)} \big] 
        + F^{--}(\bn_a; \bx_a) \, \text{Re} \big[ \hat{A}_-^2 (\bx_a) \, e^{-2im (t - L_a)} \big] 
        \notag \\
        &\qquad\quad + F^{+-}(\bn_a; \bx_a) \, \text{Re} \big[ \hat{A}_+ (\bx_a) \, \hat{A}_- (\bx_a) \, e^{-2im (t - L_a)} \big] 
    \Big\} \,.
    \label{eqzP1}
\end{align}
Here, we defined the pattern functions $F^{++}(\bn; \bx)$, $F^{--}(\bn; \bx)$, and $F^{+-}(\bn; \bx)$ at a point $\bx$ as 
\begin{align}
    F^{++}(\bn; \bx) &\equiv \frac{1}{2} \big[ 1 - 4 (\bn \cdot \be^+(\bx))^2 \big] \,, 
    \label{eqF++1}\\
    F^{--}(\bn; \bx) &\equiv \frac{1}{2} \big[ 1 - 4 (\bn \cdot \be^-(\bx))^2 \big] \,, 
    \label{eqF--1}\\
    F^{+-}(\bn; \bx) &\equiv -4 (\bn \cdot \be^+(\bx)) (\bn \cdot \be^-(\bx)) \,, 
    \label{eqF+-1}
\end{align}
with $\bn \cdot \be^P(\bx) \equiv n^j e^P_j(\bx)$ for $P = +, -$.
The dependence of the pattern functions on the direction $\mathbf{n}$ is shown in Fig.~\ref{fig:pattern}.
These functions characterize how much the pulses from the pulsar in the direction $\bn$ undergo the redshift by the vector DM at the point $\bx$.
If only the linearly polarized state is considered, without loss of generality, we can take $e^+_j(\bx)$ to be the unit vector in the direction of the vector DM so that $\hat{A}_-(\bx) = 0$.
In this case, we can reproduce the result in Refs.~\cite{Nomura:2019cvc, Omiya:2023bio}, where the linearly polarized state is studied.

Using the expressions \eqref{eqdec2} and \eqref{eqdec3}, the Earth and pulsar terms are further rewritten as 
\begin{align}
    z_{a,\text{E}}(t) &= \frac{2 \pi G \rho_{A}(\mathbf{0})}{m^2} \, 
        \big[ F^{++}(\bn_a; \mathbf{0}) \, \cos^2 \beta (\mathbf{0}) - F^{--}(\bn_a; \mathbf{0}) \, \sin^2 \beta (\mathbf{0}) \big]
        \cos (2mt - 2\alpha (\mathbf{0}))
        \notag \\
        &\quad + \frac{\pi G \rho_{A}(\mathbf{0})}{m^2} F^{+-}(\bn_a; \mathbf{0}) \, \sin 2 \beta (\mathbf{0}) \, \sin (2mt - 2 \alpha(\mathbf{0}) ) \,, 
        \label{eqzE2}\\
    z_{a,\text{P}}(t) &= \frac{2 \pi G \rho_{A}(\bx_a)}{m^2} \, 
        \big[ F^{++}(\bn_a; \bx_a) \, \cos^2 \beta (\bx_a) - F^{--}(\bn_a; \bx_a) \, \sin^2 \beta (\bx_a) \big]
        \cos (2mt - 2mL_a - 2\alpha (\bx_a))
        \notag \\
        &\quad + \frac{\pi G \rho_{A}(\bx_a)}{m^2} F^{+-}(\bn_a; \bx_a) \, \sin 2 \beta (\bx_a) \, \sin (2mt -2mL_a - 2 \alpha(\bx_a) ) \,, 
        \label{eqzP2}
\end{align}
where we used the fact that, from Eq.~\eqref{eqT00-2}, the energy density of the vector DM is expressed in terms of the field amplitude $\bar{A}$ as 
\begin{align}
    \rho_{A}(\bx) = \frac{1}{2} m^2 \bar{A}^2(\bx) \,. 
    \label{eqdensity}
\end{align}

It is conventional to define the timing residual $\Delta T_a(t)$ for the pulsar $a$ as the time integral of the redshift \cite{Detweiler:1979wn, Anholm:2008wy},
\begin{align} 
    \Delta T_a(t) \equiv \int_0^t dt' \, z_a(t') \,. 
    \label{eqR1}
\end{align}
Since the redshift $z_a(t)$ is given by the difference between the Earth term $z_{a,\text{E}}(t)$ and the pulsar term $z_{a,\text{P}}(t)$ as in Eq.~\eqref{eqz4}, it is also useful to express the timing residual as $\Delta T_a(t) = \Delta T_{a, \text{E}}(t) - \Delta T_{a, \text{P}}(t)$, where
\begin{align}
    \Delta T_{a,\text{E}}(t) \equiv \int_0^t dt' \, z_{a,\text{E}}(t') \,, 
    \qquad 
    \Delta T_{a,\text{P}}(t) \equiv \int_0^t dt' \, z_{a,\text{P}}(t') \,.
    \label{eqR2}
\end{align}
From Eqs.~\eqref{eqzE2} and \eqref{eqzP2}, we obtain the Earth and pulsar terms for the timing residual induced by the vector DM as 
\begin{align}
    \Delta T^{\text{VDM}}_{a,\text{E}}(t) &= \frac{\pi G \rho_{A}(\mathbf{0})}{m^3} \big[ F^{++}(\bn_a; \mathbf{0}) \, \cos^2 \beta(\mathbf{0}) - F^{--}(\bn_a; \mathbf{0}) \, \sin^2 \beta(\mathbf{0}) \big] 
    \sin (2mt - 2\alpha(\mathbf{0})) 
    \notag \\
    &\quad - \frac{\pi G \rho_{A}(\mathbf{0})}{2m^3} F^{+-}(\bn_a; \mathbf{0}) \, \sin 2 \beta(\mathbf{0}) \, \cos (2mt - 2\alpha(\mathbf{0}))  \,,
    \label{eqRE2}\\
    \Delta T^{\text{VDM}}_{a,\text{P}}(t) &= \frac{\pi G \rho_{A}(\bx_a)}{m^3} \big[ F^{++}(\bn_a; \bx_a) \, \cos^2 \beta(\bx_a) - F^{--}(\bn_a; \bx_a) \, \sin^2 \beta(\bx_a) \big] 
    \sin (2mt - 2mL_a - 2\alpha(\bx_a)) 
    \notag \\
    &\quad - \frac{\pi G \rho_{A}(\bx_a)}{2m^3} F^{+-}(\bn_a; \bx_a) \, \sin 2 \beta(\bx_a) \, \cos (2mt - 2mL_a - 2\alpha(\bx_a))  \,,
    \label{eqRP2}
\end{align}
where only the time-dependent terms are shown, as they are observable in PTAs.
We see that the timing residual exhibits monochromatic oscillations with angular frequency $2m$.
The corresponding frequency $f$ reads 
\begin{align}
    f = \frac{m}{\pi} = 4.8 \times 10^{-9} \, \text{Hz} \, \bigg( \frac{m}{10^{-23}\,\text{eV}} \bigg)\,. 
    \label{eqfreq}
\end{align}
The magnitude of the oscillation is estimated as 
\begin{align}
    |\Delta T^{\text{VDM}}_a| \sim \frac{\pi G \rho_{A}}{m^3} \sim 3 \times 10^{-8} \, \text{sec} \, \bigg( \frac{\rho_{A}}{0.3 \, \text{GeV/cm}^3} \bigg) \bigg( \frac{10^{-23}\,\text{eV}}{m} \bigg)^3 \,,
    \label{eqR3}
\end{align}
where the local DM density $\rho_{\text{DM}} \sim 0.3 \, \text{GeV/cm}^3$ \cite{Read:2014qva} is used as the fiducial value for $\rho_{A}$.
The above estimate suggests that current PTAs with timing accuracies of $\lesssim \mu\text{sec}$ \cite{EPTA:2023sfo, NANOGrav:2023hde, Zic:2023gta, Xu:2023wog} can be sensitive to the residuals induced by ultralight DM with masses $\lesssim 3 \times 10^{-24} \,\text{eV}$, while an observation time span of $T_{\text{obs}} \gtrsim 20 \,\text{yr}$ is required to resolve the oscillation. For such light masses, PTAs can provide a means of probing or constraining the presence of DM independently of, and complementarily to, other cosmological or astrophysical observations.
In the near future, the Square Kilometre Array (SKA) project is expected to achieve even greater precision, potentially reaching timing accuracies below $100\,\text{nsec}$ \cite{Janssen:2014dka, Lommen:2015gbz}.

Let us compare the timing residuals induced by vector DM with those induced by scalar DM of the same mass $m$.
The latter is given by $\Delta T^{\text{SDM}}_{a} = \Delta T^{\text{SDM}}_{a,\text{E}} - \Delta T^{\text{SDM}}_{a,\text{P}}$ with \cite{Khmelnitsky:2013lxt}
\begin{align}
    \Delta T^{\text{SDM}}_{a,\text{E}}(t) &= \frac{\pi G \rho_\phi(\mathbf{0})}{2m^3} \sin (2mt - 2\alpha(\mathbf{0})) \,,
    \label{eqscaRE1}\\
    \Delta T^{\text{SDM}}_{a,\text{P}}(t) &= \frac{\pi G \rho_\phi(\bx_a)}{2m^3} \sin (2mt - 2mL_a - 2\alpha(\bx_a)) \,,
    \label{eqscaRE2}
\end{align}
where $\rho_{\phi}(\bx)$ denotes the energy density of scalar DM at $\bx$, and $\alpha(\bx)$ is the phase of the oscillating scalar field at $\bx$.
Comparing Eqs.~\eqref{eqRE2} and \eqref{eqRP2} with Eqs.~\eqref{eqscaRE1} and \eqref{eqscaRE2}, it is clear that the residuals induced by vector DM have a distinctive directional dependence on the pulsar's direction $\bn_a$, which is absent in the scalar case.
Furthermore, the directional dependence varies depending on the angle $\beta$, which characterizes the polarization of the vector DM.

\section{Angular correlation}
\label{sec:angular_cor}

\subsection{Angular correlation produced by vector DM}

As used in gravitational wave detection by interferometers, taking correlations between different detectors is an effective way for extracting signals from noisy data \cite{Michelson:1987nae,Christensen:1992wi,Flanagan:1993ix}.
In the detection of gravitational wave background with PTA experiments, it is crucial to verify that the angular correlations between the timing residuals of different pulsars exhibit the characteristic pattern known as the Hellings--Downs curve \cite{Hellings:1983fr, Allen:2023kib}.
In this section, we compute the angular correlation of the timing residuals induced by vector DM. The results will be useful in distinguishing vector DM contributions from other contributions, including gravitational waves, in the PTA analysis.

We consider the following cross-correlation between the timing residuals of two pulsars induced by vector DM, 
\begin{align}
    C_{\text{VDM}}(\tau; \xi) \equiv \braket{ \Delta T^{\text{VDM}}_a(t) \, \Delta T^{\text{VDM}}_b(t+\tau)} - \braket{ \Delta T^{\text{VDM}}_a(t) } \braket{ \Delta T^{\text{VDM}}_b(t+\tau) } \,,
    \label{cor1}
\end{align}
where $\tau$ is a parameter of time difference.
The angle brackets $\braket{\cdots}$ represent two averages: 1) a time average over the observation period $T_{\text{obs}} \sim 10^{1}$--$10^2 \, \text{yr}$, and 2) an average over pulsar pairs $(a,b)$ separated by a fixed angle $\xi$ on the celestial sphere.
In the following, we assume that the pulsars are uniformly distributed on the sphere for simplicity.
For an isotropic gravitational wave background, averaging over pulsar pairs is equivalent to averaging over the propagation directions of the gravitational wave for a fixed pulsar pair. In our case, however, the background vector DM defines particular directions, so we adopt the more practical procedure of averaging over pulsar pairs.
This is analogous to the pulsar correlation analysis for deterministic and anisotropic gravitational waves, such as those produced by isolated binary sources \cite{Cornish:2013aba}.

Noting that the timing residual is expressed as $\Delta T^{\text{VDM}}_a(t) = \Delta T^{\text{VDM}}_{a,\text{E}}(t) - \Delta T^{\text{VDM}}_{a,\text{P}}(t)$, we obtain 
\begin{align}
    \braket{ \Delta T^{\text{VDM}}_a(t) \, \Delta T^{\text{VDM}}_b(t+\tau)}
    &= \braket{ \Delta T^{\text{VDM}}_{a,\text{E}}(t) \, \Delta T^{\text{VDM}}_{b,\text{E}}(t+\tau) }
     - \braket{ \Delta T^{\text{VDM}}_{a,\text{P}}(t) \, \Delta T^{\text{VDM}}_{b,\text{E}}(t+\tau) }
     \notag \\
     &\quad - \braket{ \Delta T^{\text{VDM}}_{a,\text{E}}(t) \, \Delta T^{\text{VDM}}_{b,\text{P}}(t+\tau) }
     + \braket{ \Delta T^{\text{VDM}}_{a,\text{P}}(t) \, \Delta T^{\text{VDM}}_{b,\text{P}}(t+\tau) } \,.
     \label{cor2}
\end{align}
Recall that the pulsar term $\Delta T^{\text{VDM}}_{a, \text{P}}(t)$ in Eq.~\eqref{eqRP2} contains a phase in the form of $2mL_a$, that depends on the pulsar distance $L_a$.
Given that the typical pulsar distances satisfy $L_a \gtrsim 100 \, \text{pc}$, we find $m L_a \gtrsim 10^2$ for $m = 10^{-23} \, \text{eV}$, indicating that the pulsar term $\Delta T^{\text{VDM}}_{a, \text{P}}(t)$ oscillates rapidly with respect to $L_a$. 
Therefore, after averaging over many pulsars, the contributions from the last three terms in Eq.~\eqref{cor2}, which involve pulsar terms, become suppressed relative to the first term.
Furthermore, noting that the observation period $T_{\text{obs}} \sim 10^{1}$--$10^2 \, \text{yr}$ is comparable to or longer than the oscillation period $f^{-1} = \pi/m \sim 7\,\text{yr}\,(10^{-23}\,\text{eV}/m)$, the contributions from the oscillating terms such as $\cos (2mt)$ and $\sin (2mt)$ become suppressed on the time average.
Consequently, the cross-correlation \eqref{cor1} is approximated as 
\begin{align}
    C_{\text{VDM}}(\tau; \xi) &\approx \braket{ \Delta T^{\text{VDM}}_{a,\text{E}}(t) \, \Delta T^{\text{VDM}}_{b,\text{E}}(t+\tau) }
    \notag \\
    &\approx \bigg( \frac{\pi G \rho_{A}(\mathbf{0})}{m^3} \bigg)^2 
    \bigg\{ 
        \frac{1}{2} \cos (2m\tau) 
        \Big[ 
            \braket{ F^{++}(\bn_a; \mathbf{0}) \, F^{++}(\bn_b; \mathbf{0}) } \cos^4 \beta(\mathbf{0}) 
            + \braket{ F^{--}(\bn_a; \mathbf{0}) \, F^{--}(\bn_b; \mathbf{0}) } \sin^4 \beta(\mathbf{0})     
            \notag \\
    &\qquad
            - \big( \braket{ F^{++}(\bn_a; \mathbf{0}) \, F^{--}(\bn_b; \mathbf{0}) } + \braket{ F^{--}(\bn_a; \mathbf{0}) \, F^{++}(\bn_b; \mathbf{0}) } - \braket{ F^{+-}(\bn_a; \mathbf{0}) \, F^{+-}(\bn_b; \mathbf{0}) } \big) \sin^2 \beta(\mathbf{0}) \cos^2 \beta(\mathbf{0}) 
        \Big]
        \notag \\
    &\quad + \frac{1}{2} \sin (2m\tau) 
        \Big[ 
            \big( \braket{ F^{++}(\bn_a; \mathbf{0}) \, F^{+-}(\bn_b; \mathbf{0}) } - \braket{ F^{+-}(\bn_a; \mathbf{0}) \, F^{++}(\bn_b; \mathbf{0}) } \big) \sin \beta(\mathbf{0}) \cos^3 \beta(\mathbf{0}) 
            \notag \\
    &\qquad
            + \big( \braket{ F^{+-}(\bn_a; \mathbf{0}) \, F^{--}(\bn_b; \mathbf{0}) } - \braket{ F^{--}(\bn_a; \mathbf{0}) \, F^{+-}(\bn_b; \mathbf{0}) } \big) \sin^3 \beta(\mathbf{0}) \cos \beta(\mathbf{0}) 
        \Big]
    \bigg\}\,.
    \label{cor3}
\end{align}
In the second equality, the time average is performed, where we assume that quantities associated with the vector DM, such as $\rho_{A}(\mathbf{0})$ and $\beta(\mathbf{0})$, remain constant over the observation period $T_{\text{obs}}$.
This assumption is justified by the fact that the coherence time \eqref{tau_coh} is much longer than $T_{\text{obs}}$.
The basis vectors $\be^+(\mathbf{0})$ and $\be^-(\mathbf{0})$, which define the principal axes of the polarization ellipse at the Earth, are also considered as constant.
Hence, in Eq.~\eqref{cor3}, the angle brackets for the product of the pattern functions, such as $\braket{ F^{++}(\bn_a; \mathbf{0}) \, F^{++}(\bn_b; \mathbf{0}) }$, contain only the average over the pulsar pairs.
This average is performed in Appendix \ref{app:pulsar_average}.
As a result, the cross-correlation is found to be 
\begin{align}
    C_{\text{VDM}}(\tau; \xi) &\approx \bigg( \frac{\pi G \rho_{A}(\mathbf{0})}{m^3} \bigg)^2 
    \bigg[ \frac{7}{60} + \frac{4}{15} \cos 2\xi + \bigg( \frac{1}{60} + \frac{2}{15} \cos 2\xi \bigg) \sin^2 2\beta(\mathbf{0}) \bigg]
    \frac{1}{2} \cos (2m\tau) 
    \notag \\
    &= \Phi_{\text{VDM}}\, \Gamma_{\text{VDM}}(\xi; \beta(\mathbf{0})) \cos (2m\tau) \,.
    \label{cor4}
\end{align}
In the last line, we introduced $\Phi_{\text{VDM}}$ and $\Gamma_{\text{VDM}}(\xi; \beta)$ as
\begin{align}
    \Phi_{\text{VDM}} &\equiv \frac{23}{60} \bigg( \frac{\pi G \rho_{A}(\mathbf{0})}{m^3} \bigg)^2 \,,
    \label{Phivector}\\
    \Gamma_{\text{VDM}}(\xi; \beta) &\equiv \frac{7}{46} + \frac{8}{23} \cos 2\xi + \bigg( \frac{1}{46} + \frac{4}{23} \cos 2\xi \bigg) \sin^2 2\beta
    \notag \\
    &= \frac{5}{138} (\cos^2 2 \beta) \, P_0(\cos \xi) + \frac{32}{69} \bigg( 1 + \frac{1}{2} \sin^2 2\beta \bigg) P_2(\cos \xi)\,,
    \label{Gammavector}
\end{align}
where $P_0(x) = 1$ and $P_2(x) = (3x^2 - 1)/2$ are Legendre polynomials.
The function $\Gamma_{\text{VDM}}(\xi; \beta)$ is normalized to satisfy $\Gamma_{\text{VDM}}(\xi=0; \beta=0) = 1/2$.
Note that the function $\Gamma_{\text{VDM}}(\xi; \beta)$ is invariant under the transformation $\beta \to - \beta$; that is, the angular correlation pattern is identical for both left- and right-handed polarizations. 
The function $\Gamma_{\text{VDM}}(\xi; \beta)$ is also invariant under the transformation $\beta \to \pi/2 - \beta$ (for $0 \leq \beta \leq \pi/2$), which can be understood from the fact that the states related by this transformation represent the same polarization state.

For comparison, we also compute the cross-correlation induced by scalar DM using Eq.~\eqref{eqscaRE1} as 
\begin{align}
    C_{\text{SDM}}(\tau; \xi) &\equiv \braket{ \Delta T^{\text{SDM}}_a(t) \, \Delta T^{\text{SDM}}_b(t+\tau)} - \braket{ \Delta T^{\text{SDM}}_a(t) } \braket{ \Delta T^{\text{SDM}}_b(t+\tau) } 
    \notag \\
    &\approx \bigg( \frac{\pi G \rho_\phi(\mathbf{0})}{2m^3} \bigg)^2 \frac{1}{2} \cos (2m\tau) 
    \notag \\
    &= \Phi_{\text{SDM}} \, \Gamma_{\text{SDM}}(\xi) \, \cos (2m\tau) \,.
    \label{corSDM}
\end{align}
In the last line, $\Phi_{\text{SDM}}$ and $\Gamma_{\text{SDM}}(\xi)$ are defined as
\begin{align}
    \Phi_{\text{SDM}} \equiv \frac{23}{60} \bigg( \frac{\pi G \rho_{\phi}(\mathbf{0})}{m^3} \bigg)^2 \,, 
    \qquad
    \Gamma_{\text{SDM}}(\xi) \equiv \frac{15}{46} \,. 
    \label{GammaSDM}
\end{align}
Here, $\Phi_{\text{SDM}}$ is defined so that $\Phi_{\text{SDM}} = \Phi_{\text{VDM}}$ is satisfied when we set $\rho_{\phi}(\mathbf{0}) = \rho_{A}(\mathbf{\mathbf{0}})$ and consider the same mass $m$. 

\begin{figure}[tb]
\centering
\includegraphics[width=0.6\textwidth]{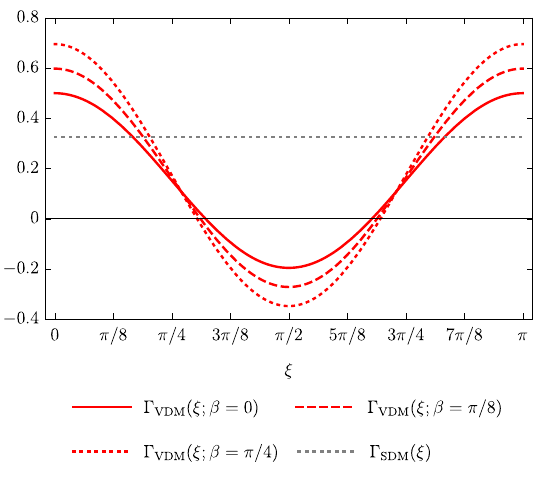}
\caption{The red curves represent the angular correlation function induced by vector DM, $\Gamma_{\text{VDM}}(\xi; \beta)$, for $\beta = 0$ (solid curve), $\beta = \pi/8$ (dashed curve), and $\beta = \pi/4$ (dotted curve). The cases $\beta = 0$ and $\beta = \pi/4$ correspond to vector DM states with perfect linear and circular polarization, respectively.
The dotted gray line represents the angular correlation function induced by scalar DM, $\Gamma_{\text{SDM}}(\xi)$.}
\label{fig:angular}
\end{figure}

In Fig.~\ref{fig:angular}, the angular correlation function induced by vector DM, $\Gamma_{\text{VDM}}(\xi;\beta)$, is plotted as a function of the pulsar separation angle $\xi$ for several values of $\beta$.
In the same figure, the angular correlation function induced by scalar DM, $\Gamma_{\text{SDM}}(\xi)$, is also plotted for comparison.
It is evident that vector DM produces the angular correlation with a nontrivial dependence on $\xi$, unlike scalar DM.
Interestingly, the quadrupolar feature of $\Gamma_{\text{VDM}}(\xi;\beta)$ is enhanced as the circular polarization increases.
In Eq.~\eqref{Gammavector}, the coefficient of the monopole $P_0(\cos \xi)$ becomes maximal, while that of the quadrupole $P_2(\cos \xi)$ becomes minimal at $\beta = 0$ and $\pm\pi/2$, corresponding to the linearly polarized state.
In contrast, the coefficient of the quadrupole $P_2(\cos \xi)$ becomes maximal, while that of the monopole $P_0(\cos \xi)$ vanishes at $\beta = \pm \pi/4$, corresponding to the state of perfect circular polarization.
The absence of the monopole component in the perfectly circularly polarized state is consistent with the fact that the trace of the stress tensor $T_{ij}$ vanishes when $\beta = \pm \pi/4$, as can be confirmed from Eqs.~\eqref{eqTjk-2} and \eqref{pol4}.
Therefore, the circular polarization of vector DM leads to a more significant difference from scalar DM in the angular correlation.

Note that the angular correlation function $\Gamma_{\text{VDM}}(\xi; \beta)$ is symmetric under the transformation $\xi \to \pi - \xi$, i.e., reflection about $\xi = \pi/2$, unlike the Hellings--Downs curve produced by stochastic gravitational waves \cite{Hellings:1983fr}. This difference originates from the fact that the vector DM forms a nearly standing wave, whereas gravitational waves are propagating at the speed of light. Indeed, the propagation of gravitational waves generates multipole modes beyond the quadrupole in the angular correlation, leading to an asymmetry under reflection about $\xi = \pi/2$ \cite{Romano:2023zhb}.

\subsection{Deformation of the Hellings--Downs curve}

In reality, the DM and gravitational wave will coexist. 
Current PTA experiments are trying to detect background gravitational waves by identifying the characteristic angular correlation known as the Hellings--Downs curve. 
If the ultralight vector field constitutes the DM, it should deform the Hellings--Downs curve.
Let us demonstrate this deformation.

We consider the cross-correlation between the timing residuals of two pulsars, which are produced by an ultralight vector DM and stochastic background gravitational waves:
\begin{align}
    C(\tau; \xi) = C_{\text{VDM}}(\tau; \xi) + C_{\text{GW}}(\tau; \xi) \,. 
    \label{cortot1}
\end{align}
Here, $C_{\text{VDM}}(\tau; \xi)$ is given by Eq.~\eqref{cor4}, and $C_{\text{GW}}(\tau; \xi)$ represents the contribution from the gravitational wave background. 
Assuming that background gravitational waves are stationary, isotropic, unpolarized, and Gaussian, the cross-correlation $C_{\text{GW}}(\tau; \xi)$ is expressed as 
\begin{align}
    C_{\text{GW}}(\tau; \xi) = \sum_{i} \Phi_{\text{GW}}(f_i) \, \Gamma_{\text{HD}}(\xi) \, \cos (2\pi f_i \tau) \,,
    \label{corGW1}
\end{align}
where we write the frequency integral as a discrete sum over frequency bins with $f_i \equiv i / T_{\text{obs}}$ determined by the total observation time $T_{\text{obs}}$, as in Ref.~\cite{NANOGrav:2023gor}.
In Eq.~\eqref{corGW1}, $\Gamma_{\text{HD}}(\xi)$ represents the Hellings--Downs curve \cite{Hellings:1983fr}, 
\begin{align}
    \Gamma_{\text{HD}}(\xi) = \frac{1}{2} - \frac{1}{4} \bigg( \frac{1 - \cos \xi}{2} \bigg) + \frac{3}{2} \bigg( \frac{1 - \cos \xi}{2} \bigg) \ln \bigg( \frac{1 - \cos \xi}{2} \bigg) \,, 
    \label{HD}
\end{align}
and $\Phi_{\text{GW}}(f_i)$ is a power spectrum density commonly parametrized as \cite{NANOGrav:2023gor} 
\begin{align}
    \Phi_{\text{GW}}(f_i) = \frac{A_{\text{GW}}^2}{12 \pi^2} \frac{1}{T_{\text{obs}}} \bigg( \frac{f_i}{f_{\text{ref}}} \bigg)^{-\gamma} f_{\text{ref}}^{-3}\,, 
    \label{PhiGW1}
\end{align}
where $A_{\text{GW}}$ is the amplitude, $\gamma$ is the spectral index, and $f_{\text{ref}}$ is a reference frequency typically set to $f_{\text{ref}} = 1\,\text{yr}^{-1}$.

From Eq.~\eqref{cor4}, the ultralight vector DM contributes to the cross-correlation in a single frequency bin that includes $f = m/\pi$. 
We can write the cross-correlation in the frequency bin $f = m/\pi$ as 
\begin{align}
    C(\tau; \xi)_{f=m/\pi} = \Phi_{\text{eff}}(m/\pi) \, \Gamma_{\text{eff}}(\xi) \, \cos(2m\tau) \,,
    \label{cortot2}
\end{align}
where we defined the effective power spectral density $\Phi_{\text{eff}}(m/\pi)$ and the angular correlation curve $\Gamma_{\text{eff}}(\xi)$ as 
\begin{align}
    \Phi_{\text{eff}}(m/\pi)
    &\equiv \Phi_{\text{GW}}(m/\pi) 
    \bigg( \frac{\Phi_{\text{VDM}}}{\Phi_{\text{GW}}(m/\pi)} 2\Gamma_{\text{VDM}}(\xi=0; \beta) + 1 \bigg) \,,
    \label{cortot3}\\
    \Gamma_{\text{eff}}(\xi) &\equiv \frac{\Phi_{\text{GW}}(m/\pi)}{\Phi_{\text{VDM}} \,  2 \Gamma_{\text{VDM}}(\xi=0;\beta) + \Phi_{\text{GW}}(m/\pi)} 
    \bigg( \frac{\Phi_{\text{VDM}}}{\Phi_{\text{GW}}(m/\pi)} \Gamma_{\text{VDM}}(\xi; \beta) + \Gamma_{\text{HD}}(\xi) \bigg) \,.
    \label{cortot4}
\end{align}
Here, we write $\beta(\mathbf{0})$ simply as $\beta$, and normalize $\Gamma_{\text{eff}}(\xi)$ such that $\Gamma_{\text{eff}}(\xi=0) = 1/2$.
Note that $\Gamma_{\text{VDM}}(\xi=0; \beta) = 1/2 + (9/46)\sin^2 2 \beta$ from Eq.~\eqref{Gammavector}.

As a demonstration, let us employ the power spectral density of gravitational waves inferred from the NANOGrav 15 yr observations; $A_{\text{GW}} = 2.4 \times 10^{-15}$ at $\gamma = 13/3$ with $T_{\text{obs}} \sim 15\, \text{yr}$ and $f_{\text{ref}} = 1 \, \text{yr}^{-1}$ \cite{NANOGrav:2023gor}. This leads to
\begin{align}
    \Phi_{\text{GW}}(m/\pi) 
    \sim 1 \times 10^{-14} \, \text{sec}^2 \, \bigg( \frac{m}{10^{-23} \, \text{eV}} \bigg)^{-13/3} \bigg( \frac{15 \, \text{yr}}{T_{\text{obs}}} \bigg)\,. 
    \label{PhiGW2}
\end{align}
In contrast, from Eq.~\eqref{Phivector}, we estimate $\Phi_{\text{VDM}}$ as
\begin{align}
    \Phi_{\text{VDM}} \sim 
    4 \times 10^{-16}\, \text{sec}^2 \, \bigg( \frac{10^{-23}\, \text{eV}}{m} \bigg)^6 \bigg( \frac{\rho_{A}(\mathbf{0})}{0.3 \, \text{GeV/cm}^3} \bigg)^2 \,.
    \label{Phivector2}
\end{align}
When $\Phi_{\text{VDM}}$ is non-negligible compared to $\Phi_{\text{GW}}(m/\pi)$, the effective spectral density $\Phi_{\text{eff}}(m/\pi)$ deviates from $\Phi_{\text{GW}}(m/\pi)$. In this case, the vector DM induces an excess in the power spectrum at the frequency bin corresponding to $f=m/\pi$. Note that this also holds for scalar DM with $\Phi_{\text{SDM}}$ given by Eq.~\eqref{GammaSDM}.

\begin{figure}[tb]
\centering
\includegraphics[width=\textwidth]{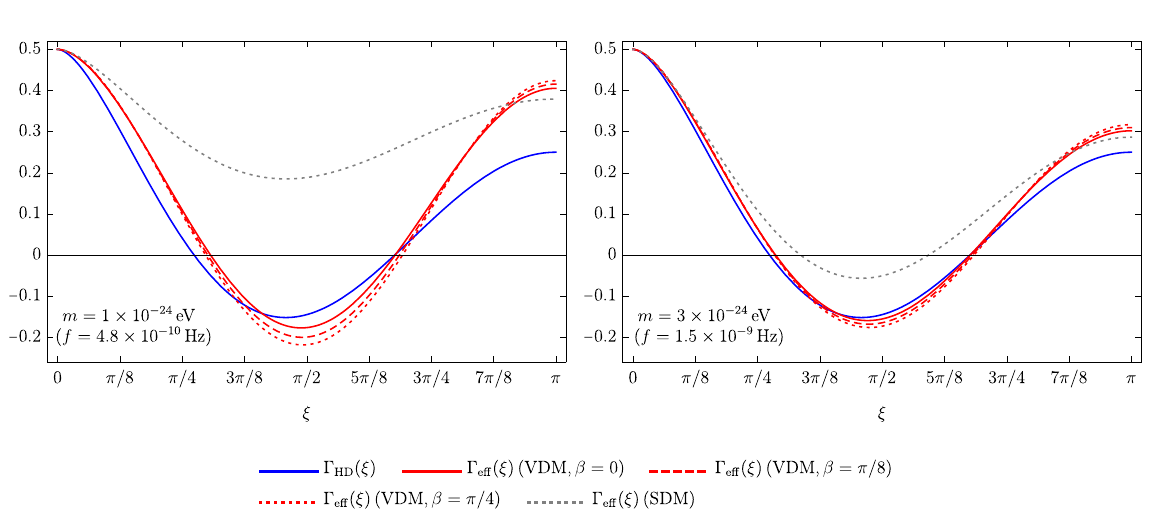}
\caption{The blue curves show the Hellings--Downs curve $\Gamma_{\text{HD}}(\xi)$. The red curves represent the effective angular correlation curve $\Gamma_{\text{eff}}(\xi)$
modified by ultralight vector DM with $\beta = 0$ (solid curves), $\beta = \pi/8$ (dashed curves), and $\beta = \pi/4$ (dotted curves). The gray dotted curves show the modification due to ultralight scalar DM. The DM mass is set to $m = 1 \times 10^{-24}\,\text{eV}$ in the left panel and $m = 3 \times 10^{-24}\,\text{eV}$ in the right panel, corresponding to the frequency $f = 4.8 \times 10^{-10}\,\text{Hz}$ and $f = 1.5 \times 10^{-9}\,\text{Hz}$, respectively.
We use $\rho_{A}(\text{or}\,\rho_\phi) = 0.3 \,\text{GeV/cm}^3$ and $\Phi_{\text{GW}}$ given by Eq.~\eqref{PhiGW2} with $T_{\text{obs}} = 15\,\text{yr}$.}
\label{fig:Gammaeff}
\end{figure}

Furthermore, in the presence of ultralight DM, the angular correlation curve can be deformed from the Hellings--Downs curve $\Gamma_{\text{HD}}(\xi)$ to the effective one $\Gamma_{\text{eff}}(\xi)$ given in Eq.~\eqref{cortot4} at the frequency $f=m/\pi$.
In Fig.~\ref{fig:Gammaeff}, we plot $\Gamma_{\text{eff}}(\xi)$ modified by the vector DM as red curves for masses $m = 1 \times 10^{-24}\, \text{eV}$ and $3 \times 10^{-24}\, \text{eV}$, assuming $\rho_{A} = 0.3 \, \text{GeV/cm}^3$.
For a fixed density $\rho_{A}$, the deformation becomes more significant as the mass $m$ decreases.
Additionally, for a fixed mass, the deformation near $\xi = \pi/2$ is enhanced for larger circular polarization of the vector DM.
Specifically, $\Gamma_{\text{eff}}(\xi)$ becomes more concave as the angle $\beta$ approaches $\pm \pi/4$.
This can be understood from the structure of $\Gamma_{\text{VDM}}(\xi;\beta)$ in Eq.~\eqref{Gammavector}, where the quadrupole component becomes stronger while the monopole component becomes weaker as $\beta \to \pm \pi/4$. 
In Fig.~\ref{fig:Gammaeff}, for comparison, we also plot the curve $\Gamma_{\text{eff}}(\xi)$ modified by scalar DM as a gray dashed curve, which is obtained by replacing $\Phi_{\text{VDM}}$ with $\Phi_{\text{SDM}}$ and $\Gamma_{\text{VDM}}$ with $\Gamma_{\text{SDM}}$ in Eq.~\eqref{cortot4}.
It is obvious that the scalar DM makes the angular correlation curve less bent, since $\Gamma_{\text{SDM}}$ contains only the monopole ($\xi$-independent) component.

From the perspective of ultralight DM searches, the gravitational wave background acts as correlated noise. Moreover, if the background is anisotropic, it may generate angular correlations that are strongly degenerate with the DM signal. However, the distinct frequency dependence, namely, a monochromatic power spectrum for DM versus a broadband spectrum for the gravitational wave background, can be exploited to extract the DM contribution.
Specifically, for a given target DM mass (frequency) bin, the noise spectrum can be estimated from the surrounding frequency bins and subtracted. 
If the DM is present in that bin, the residual component should then exhibit the angular correlation pattern derived in this work.

In Fig.~\ref{fig:Gammaeff}, we present only the cases of $m = 1 \times 10^{-24} \,\text{eV}$ and $3 \times 10^{-24} \,\text{eV}$ as illustrative examples, in which the deformation of the angular correlation is visually evident.
For larger DM masses, the deformation becomes less discernible by eye; however, in principle, the DM abundance can still be inferred through high-precision pulsar timing observations.
In Appendix \ref{app:estimation}, we discuss the expected accuracy in estimating the DM abundance using angular correlations, given the number of pulsars and the level of timing residual noise.

\section{Conclusion}
\label{sec:conclusion}

In this paper, we study observational signatures of ultralight vector DM in PTA experiments, allowing various polarization states of the vector field. 
The present study corresponds to the generalization of previous works \cite{Nomura:2019cvc, Omiya:2023bio}, where only the linearly polarized state is considered. 
Ultralight vector DM is supposed to oscillate at an angular frequency given by its mass $m$, and it has anisotropic stress oscillating at an angular frequency of $2m$.
The oscillating stress induces perturbations of the spacetime metric, which can modify the arrival times of pulses emitted from pulsars.
We derive the expressions of the pulsar timing residuals induced by the vector DM in Eqs.~\eqref{eqRE2} and \eqref{eqRP2}. These also oscillate with the angular frequency of $2m$ and exhibit nontrivial directional dependence expressed by pattern functions $F^{++}$, $F^{--}$, and $F^{+-}$. The directional dependence reflects the anisotropic property specific to vector DM, unlike scalar DM. Furthermore, the contribution of each pattern function depends on the angle $\beta$, which characterizes the ellipticity of the polarization of the vector DM. 

We also calculate cross-correlations between the timing residuals of different pulsars induced by the vector DM, which will be an additional contribution to the cross-correlations created by background gravitational waves.
Since the DM-induced signal is monochromatic at the specific frequency $f = m/\pi$, an excess of the power spectrum will appear at that frequency if the contribution of the DM is comparable to or greater than that of gravitational waves.
Furthermore, as shown in Fig.~\ref{fig:angular}, we identify the angular correlation curve produced by the vector DM in various polarization states.
Interestingly, the monopole component of the angular correlation is reduced and the quadrupole component is enhanced as the state of the vector DM changes from linear $(\beta = 0, \pm \pi/2)$ to circular $(\beta = \pm \pi/4)$ polarization.
In particular, the perfectly circularly polarized vector DM exhibits a completely quadrupole angular correlation.
As a result, the circularly polarized vector DM deforms the Hellings--Downs curve to be more bent. This is the opposite trend from the deformation caused by scalar DM, which has only a monopole component.
As shown in Fig.~\ref{fig:Gammaeff}, for a given DM density, the deformation of the Hellings--Downs curve is more significant for DM with smaller masses.
The angular correlation curves derived here will be beneficial to distinguish between the contribution of gravitational waves and that of ultralight DM, and also to determine whether the DM is a scalar or vector field.
Although recent Lyman-$\alpha$ observations suggest constraints on the abundance of ultralight DM in a mass range $m \lesssim 10^{-21}\, \text{eV}$ \cite{Irsic:2017yje, Kobayashi:2017jcf}, and observations of giant galaxies can test the fuzzy DM scenario with $m \sim 10^{-22} \, \text{eV}$ \cite{DeLaurentis:2022nrv}, PTA observations offer an independent means of constraining or searching for DM on the galactic scale.
In Fig.~\ref{fig:Gammaeff}, we show the cases of $m=1 \times 10^{-24}\, \text{eV}$ and $3 \times 10^{-24}\,\text{eV}$ as illustrative examples in which the deformation of the angular correlation is clearly visible.
Even for larger masses, the DM abundance can, in principle, be estimated or constrained from pulsar observations, provided that a sufficiently large number of pulsars are observed with low timing noise.

In this study, we focus on the spin-1 vector DM. It would be interesting to extend the present analysis to higher spin fields \cite{Aoki:2016zgp,Aoki:2017cnz,Babichev:2016bxi,Marzola:2017lbt,Manita:2022tkl,Armaleo:2020yml,Wu:2023dnp}, as they can have richer polarization states.
Furthermore, a similar analysis could be performed to identify the effects in astrometry \cite{Book:2010pf}. 
We leave these studies for future work.

\section*{Acknowledgments}
K.N. was supported by JSPS KAKENHI Grant Numbers JP24KJ0117 and JP25K17389. H. O.~was supported by JSPS KAKENHI Grant Numbers JP23H00110 and JP25K17388, and Yamada Science Foundation. T. T. was supported by the JSPS KAKENHI Grant Numbers JP23H00110, 24H00963 and 24H01809.

\appendix

\section{Noether currents}
\label{app:Noether}

In this Appendix, we consider a massive vector field in a flat space and study the Noether currents.
We consider the following Lagrangian density,
\begin{align}
    \mathcal{L} &= - \frac{1}{4} \eta^{\mu\rho} \eta^{\nu\sigma} F_{\mu\nu} F_{\rho\sigma} - \frac{1}{2} m^2 \eta^{\mu\nu} A_\mu A_\nu \,,
    \label{eqLag1}
\end{align}
where $\eta_{\mu\nu}$ is the Minkowski metric.
The Noether current associated with the translation invariance is the canonical energy-momentum tensor, which we denote as $T_{\text{c,} \mu \nu}$. Following the conventional definition, we have
\begin{align}
    {{T_{\text{c}}}^\mu}_\nu &\equiv - \frac{\pd \mathcal{L}}{\pd (\pd_\mu A_\lambda)} \pd_\nu A_\lambda + \delta^\mu_\nu \mathcal{L} 
    \notag \\
    &= F^{\mu\lambda} \pd_\nu A_\lambda + \delta^\mu_\nu \bigg( - \frac{1}{4} F^{\rho\lambda} F_{\rho\lambda} - \frac{1}{2} m^2 A^\lambda A_\lambda \bigg) \,.
    \label{eqTc1}
\end{align}
We can explicitly show that this tensor indeed satisfies the conservation law, $\pd_\mu {{T_{\text{c}}}^\mu}_\nu = 0$, by using the equation of motion $\pd_\nu F^{\nu\mu} = m^2 A^\mu$.
However, this tensor is not symmetric in the indices, ${T_{\text{c,}\mu\nu}} \neq {T_{\text{c,}\nu\mu}}$. 
The symmetrized version is expected to be given by the energy-momentum tensor defined through the variation with respect to the metric, which we denote as $T_{\mu\nu}$ and derive in Eq.~\eqref{eqEMT1}.
Indeed, the difference between them turns out to be  
\begin{align}
    {T^\mu}_\nu - {{T_{\text{c}}}^\mu}_\nu 
    &= - F^{\mu\lambda} \pd_\lambda A_\nu + m^2 A^\mu A_\nu 
    = \pd_\lambda (F^{\lambda \mu} A_\nu) \,,
    \label{eqTc2}
\end{align}
where we used the equation of motion in the latter equality.
From the identity $\pd_\mu \pd_\lambda (F^{\lambda \mu} A_\nu) = 0$, it is obvious that the addition of $\pd_\lambda (F^{\lambda \mu} A_\nu)$ does not spoil the conservation law.

We also obtain the Noether current associated with the Lorentz invariance, which we denote as ${\mathcal{M}^\mu}_{\sigma \rho}$ and call the (four-dimensional) angular momentum density.
This consists of the spin part ${\mathcal{S}^\mu}_{\sigma \rho}$ and the orbital part ${\mathcal{L}^\mu}_{\sigma \rho}$\,,
\begin{align}
    {\mathcal{M}^\mu}_{\sigma \rho} &= {\mathcal{S}^\mu}_{\sigma \rho} + {\mathcal{L}^\mu}_{\sigma \rho}\,. 
    \label{eqangM1}
\end{align}
The spin part is defined by 
\begin{align}
    {\mathcal{S}^\mu}_{\sigma \rho} &\equiv i\, \frac{\pd \mathcal{L}}{\pd (\pd_\mu A_\lambda)} {(S_{\rho\sigma})_\lambda}^\nu A_\nu \,, 
    \qquad
    \text{with} \quad {(S_{\rho\sigma})_\lambda}^\nu \equiv -i (\eta_{\lambda\rho} \delta^\nu_\sigma - \eta_{\lambda\sigma} \delta^\nu_\rho) \,.
    \label{eqspinS1}
\end{align}
The orbital part is defined by 
\begin{align}
    {\mathcal{L}^\mu}_{\sigma \rho} &\equiv x_\sigma {{T_{\text{c}}}^\mu}_\rho - x_\rho {{T_{\text{c}}}^\mu}_\sigma \,.
    \label{eqorbL1}
\end{align}
Explicitly, the spin part is given by ${\mathcal{S}^\mu}_{\sigma \rho} = A_\rho {F^{\mu}}_{\sigma} - A_\sigma {F^\mu}_\rho$.

Now, let us consider the configuration of the vector field described in Sec.~\ref{sec:vectorDM}. 
There, the vector field is represented as Eq.~\eqref{eqvdmbg1} with nearly uniform amplitudes and phases.
In this case, neglecting the spatial gradient of the field, we see that the orbital angular momentum density ${\mathcal{L}^\mu}_{ij}$ trivially satisfies the conservation law, $\partial_\mu {\mathcal{L}^\mu}_{ij} = 0$.
Hence, the spin density ${\mathcal{S}^\mu}_{ij}$ must be solely conserved, $\partial_\mu {\mathcal{S}^\mu}_{ij} = 0$. 
Noting that we have ${\mathcal{S}^0}_{ij} = A_i \dot{A}_j - A_j \dot{A}_i$ and ${\mathcal{S}^k}_{ij} = 0$ neglecting the spatial gradient of the field, the conservation law leads to ${\dot{\mathcal{S}}^0}_{~ij} = 0$. 
Equivalently, it is expressed as $d\mathbf{S}/dt = 0$, where $\mathbf{S}=(S^i)$ is the three-dimensional spin vector defined by $S^i \equiv (1/2) \varepsilon^{ijk} {\mathcal{S}^0}_{jk}$ with the anti-symmetric tensor $\varepsilon^{ijk}$.
Using ${\mathcal{S}^0}_{ij} = A_i \dot{A}_j - A_j \dot{A}_i$, the spin vector $\mathbf{S}$ is given by
\begin{align}
    \mathbf{S} = \mathbf{A} \times \dot{\mathbf{A}} \,. 
    \label{eqspinS3}
\end{align}
Since the vector $\mathbf{A}$ is orthogonal to $\mathbf{S}$, the conservation of $\mathbf{S}$ implies that the vector $\mathbf{A}$ remains confined to a single two-dimensional plane.

\section{Polarized state of a vector field}
\label{app:ellipse}

In this Appendix, we describe an arbitrarily elliptically polarized state of a vector field following Sec.~2.4 of the textbook \cite{rybicki2024radiative}. 
From the conservation of angular momentum, the vector at a given point $\bx$ remains confined to a single plane. 
In this plane, we take the $+$ and $-$ axes along the principal axes of the ellipse, as shown in Fig.~\ref{fig:ellipse1}.
In this frame, an elliptically polarized vector with angular frequency $m$ can be expressed as 
\begin{align}
    (A_{+}, A_{-}) &= (\bar{A} \cos \beta \cos (mt') \,, \bar{A} \sin \beta \sin (mt') ) \,,
    \label{eqvdmbg3}
\end{align}
where $\bar{A}$ and $\beta$ are constants with $\bar{A} > 0$ and $-\pi/2 \leq \beta \leq \pi/2$.
The lengths of the principal axes are $\bar{A} \,|\cos \beta|$ and $\bar{A} \, |\sin \beta|$.
The time coordinate $t'$ is chosen so that the vector points along the $+$ axis at $t'=0$.
In Eq.~\eqref{pol2}, the time coordinate $t$ is used, which is related to $t'$ through $mt' = mt - \alpha$ with an arbitrary phase $\alpha$.

Let us see that the elliptically polarized state \eqref{eqvdmbg3} (or \eqref{pol2}) can be rewritten in the form of Eq.~\eqref{pol1} through a coordinate transformation.
We define the new $x$ and $y$ axes by rotating the $+$ and $-$ axes by an arbitrary angle $\chi$. Then, the vector components in the new frame read
\begin{align}
    A_x &= A_{+} \cos \chi + A_{-} \sin \chi 
    \notag \\
    &= \bar{A} \, [\cos \beta \cos \chi \cos (mt') + \sin \beta \sin \chi \sin (mt') ] \,, 
    \label{eqAx1}\\
    A_y &= - A_{+} \sin \chi + A_{-} \cos \chi 
    \notag \\
    &= \bar{A} \, [ - \cos \beta \sin \chi \cos (mt') + \sin \beta \cos \chi \sin (mt') ] \,. 
    \label{eqAy1}
\end{align}
Now let us introduce $\bar{A}_x$, $\alpha_x'$, $\bar{A}_y$, and $\alpha_y'$ as follows,
\begin{align}
    \bar{A}_x &\equiv \bar{A} \sqrt{ \cos^2 \beta \cos^2 \chi + \sin^2 \beta \sin^2 \chi } \,, 
    \label{eqAx2}\\
    \tan \alpha_x' &\equiv \frac{\sin \beta \sin \chi}{\cos \beta \cos \chi} \,,
    \label{eqAx3}\\
    \bar{A}_y &\equiv \bar{A} \sqrt{ \cos^2 \beta \sin^2 \chi + \sin^2 \beta \cos^2 \chi } \,, 
    \label{eqAy2}\\
    \tan \alpha_y' &\equiv - \frac{\sin \beta \cos \chi}{\cos \beta \sin \chi} \,.
    \label{eqAy3}
\end{align}
Using these relations, the vector components in the new frame \eqref{eqAx1} and \eqref{eqAy1} are represented as
\begin{align}
    (A_x, A_y) &= (\bar{A}_x \cos(mt' - \alpha_x'), \bar{A}_y \cos(mt' - \alpha_y') ) \,.
    \label{eqvdmbg2}
\end{align}
Finally, we can define new phases $\alpha_x$ and $\alpha_y$ as $\alpha_x \equiv \alpha_x' + \alpha$ and $\alpha_y \equiv \alpha_y' + \alpha$ to reproduce the expression \eqref{pol1}.

\section{Pulsar average}
\label{app:pulsar_average}

In this Appendix, we calculate the averages of products of the pattern functions over pulsar pairs, such as $\braket{ F^{++}(\bn_a; \mathbf{0}) \, F^{++}(\bn_b; \mathbf{0}) }$, which appear in Eq.~\eqref{cor3}.
These averages are taken over the directions of the two pulsars, $\bn_a$ and $\bn_b$, fixing their angular separation as $\bn_a \cdot \bn_b = \cos\xi$.
First, we note that, for a given pulsar pair with a separation angle $\xi$, we can perform the coordinate transformation so that the unit vectors $\bn_a = (n_a^j)$ and $\bn_b = (n_b^j)$ are represented as 
\begin{align}
    n_a^j &= (0,0,1) \,, 
    \label{eqna1}\\
    n_b^j &= (\sin \xi, 0, \cos \xi) \,.
    \label{eqnb2}
\end{align}
We parametrize the components of the orthonormal vectors $\be^+(\mathbf{0})$ and $\be^-(\mathbf{0})$ in this frame as 
\begin{align}
    e^+_j(\bm{0}) 
    &= (\sin \theta \cos \phi, \sin \theta \sin \phi, \cos \theta) \,,
    \label{eqe+1}\\
    e^-_j(\bm{0}) 
    &= (- \sin \psi \sin \phi + \cos \psi \cos \theta \cos \phi, \sin \psi \cos \phi + \cos \psi \cos \theta \sin \phi, -\cos\psi \sin \theta) \,.
    \label{eqe-1}
\end{align}
where $\theta$, $\phi$, and $\psi$ are defined within $0 \leq \theta \leq \pi$, $-\pi \leq \phi < \pi$, and $-\pi \leq \psi < \pi$, respectively. 
Using these parametrizations, the pattern functions \eqref{eqF++1}--\eqref{eqF+-1} at the Earth ($\bx = \mathbf{0}$) are expressed as 
\begin{align}
    F^{++}(\bn_a; \mathbf{0}) &= \frac{1}{2} - 2\cos^2 \theta \,, 
    \label{eqF++a1}\\
    F^{--}(\bn_a; \mathbf{0}) &= \frac{1}{2} - 2 \cos^2\psi \sin^2 \theta \,, 
    \label{eqF--a1}\\
    F^{+-}(\bn_a; \mathbf{0}) &= 4 \cos \psi \cos \theta \sin \theta \,, 
    \label{eqF+-a1}\\
    F^{++}(\bn_b; \mathbf{0}) &= \frac{1}{2} - 2 (\sin \xi \sin \theta \cos \phi + \cos \xi \cos \theta)^2  \,,
    \label{eqF++b1}\\
    F^{--}(\bn_b; \mathbf{0}) &= \frac{1}{2} - 2 ( \sin \xi \cos \psi \cos \theta \cos \phi - \sin \xi \sin \psi \sin \phi - \cos \xi \cos \psi \sin \theta )^2 \,,
    \label{eqF--b1}\\
    F^{+-}(\bn_b; \mathbf{0}) &= -4 (\sin \xi \sin \theta \cos \phi + \cos \xi \cos \theta) ( \sin \xi \cos \psi \cos \theta \cos \phi - \sin \xi \sin \psi \sin \phi - \cos \xi \cos \psi \sin \theta ) \,.
    \label{eqF+-b1}
\end{align}
The above coordinate transformation can be performed for any pulsar pair with the separation angle $\xi$. This indicates that, assuming that the pulsars are uniformly distributed on the celestial sphere, the average over the pulsar directions can be computed as the average over the angles $\theta$, $\phi$, and $\psi$ in the above frame.
For example, the average $\braket{ F^{++}(\bn_a; \mathbf{0}) \, F^{++}(\bn_b; \mathbf{0}) }$ is computed as 
\begin{align}
    \braket{ F^{++}(\bn_a; \mathbf{0}) \, F^{++}(\bn_b; \mathbf{0}) }
    &= \frac{1}{8\pi^2} \int_0^\pi d\theta \, \sin \theta \int_{-\pi}^{\pi} d\phi \int_{-\pi}^{\pi} d\psi \, \bigg( \frac{1}{2} - 2\cos^2 \theta \bigg) \bigg( \frac{1}{2} - 2 (\sin \xi \sin \theta \cos \phi + \cos \xi \cos \theta)^2 \bigg) 
    \notag \\
    &= \frac{7}{60} + \frac{4}{15} \cos 2\xi \,. 
    \label{FppaFppb}
\end{align}
Similarly, we obtain 
\begin{align}
    \braket{ F^{--}(\bn_a; \mathbf{0}) \, F^{--}(\bn_b; \mathbf{0}) } 
    &= \frac{7}{60} + \frac{4}{15} \cos 2\xi \,,
    \label{FmmaFmmb}\\
    \braket{ F^{++}(\bn_a; \mathbf{0}) \, F^{--}(\bn_b; \mathbf{0}) } 
    &= \braket{ F^{--}(\bn_a; \mathbf{0}) \, F^{++}(\bn_b; \mathbf{0}) } 
    = - \frac{1}{60} - \frac{2}{15} \cos 2\xi \,,
    \label{FppaFmmb}\\
    \braket{ F^{+-}(\bn_a; \mathbf{0}) \, F^{+-}(\bn_b; \mathbf{0}) } 
    &= \frac{4}{15} + \frac{4}{5} \cos 2\xi \,, 
    \label{FpmaFpmb}\\
    \braket{ F^{++}(\bn_a; \mathbf{0}) \, F^{+-}(\bn_b; \mathbf{0}) } 
    &= \braket{ F^{+-}(\bn_a; \mathbf{0}) \, F^{++}(\bn_b; \mathbf{0}) } 
    = \braket{ F^{+-}(\bn_a; \mathbf{0}) \, F^{--}(\bn_b; \mathbf{0}) }
    = \braket{ F^{--}(\bn_a; \mathbf{0}) \, F^{+-}(\bn_b; \mathbf{0}) }
    = 0 \,.
    \label{FppaFpmb}
\end{align}

\begin{figure}[tb]
\centering
\includegraphics[width=\textwidth]{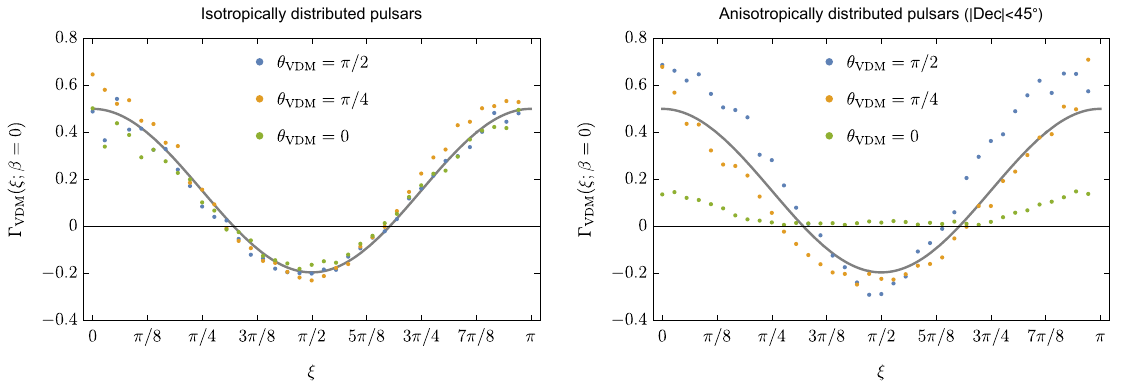}
\caption{Angular correlations induced by linearly polarized vector DM for simulated sets of 200 pulsars distributed isotropically over the sky (left) and anisotropically within the region of the sky where the declination lies within $\pm 45^\circ$ (right).
The blue, yellow, and green points correspond to vector DM orientations $\theta_{\text{VDM}} = \pi/2$, $\pi/4$, and $0$, respectively, where $\theta_{\text{VDM}}$ is measured from the north pole.
In both panels, the gray curve represents the analytical correlation assuming an isotropic pulsar distribution.}
\label{fig:AnisoDist}
\end{figure}

As done above, the calculations throughout this paper assume that pulsars are uniformly and isotropically distributed.
However, the actual distribution of pulsars exhibits anisotropy. Indeed, approximately 90\% of the pulsars in IPTA DR2 \cite{Perera:2019sca} lie within a region of the celestial sphere spanning $\pm 45$ degrees in declination.
To roughly estimate the effect of this pulsar anisotropy, we performed a simple simulation as follows.
We uniformly generated 200 pulsars within a region spanning $\pm 45$ degrees in declination. For this pulsar realization, we calculated the angular correlations induced by vector DM. For simplicity, we considered only vector DM with linear polarization ($\beta = 0$) and examined three orientations of the vector DM: $\theta_{\text{VDM}} = 0$, $\pi/4$, and $\pi/2$, where $\theta_{\text{VDM}}$ denotes the angle measured from the north pole.
The results are shown in the right panel of Figure \ref{fig:AnisoDist}, which should be compared with the isotropic case shown in the left panel. In both panels, the gray curve represents the analytical correlation predicted under the assumption of isotropic pulsar distribution.
The right panel shows that the correlation is reduced when $\theta_{\rm VDM} = 0$, i.e., when many pulsars are distributed perpendicular to the vector DM direction. On the other hand, for $\theta_{\rm VDM} > \pi/4$, the correlation approximately follows the analytical isotropic result.
This simulation is intended only as a demonstration, and a more detailed analysis using realistic pulsar distributions is left for future work.

\section{Estimation error for DM abundance}
\label{app:estimation}

In Sec.~\ref{sec:angular_cor}, we derived the angular correlation curves of timing residuals produced by vector DM. These can be used as theoretical templates for measuring the abundance of the DM. In this Appendix, we provide a rough estimate of the error in the DM abundance measurement in the presence of timing residual noise.

Let $\Delta \hat{T}_a(f_j)$ denote the observed timing residual of the pulsar $a$ in the frequency bin $f_j = j/ T_{\obs}$, where $T_{\obs}$ is the observation time span ($\sim 10^{1}$--$10^2 \, \text{yr}$).
Specifically, from the time series data $\Delta {T}_a(t)$, we define $\Delta \hat{T}_a(f_j)$ as 
\begin{align}
    \Delta \hat{T}_a(f_j) \equiv \frac{1}{T_{\obs}} \int_{-T_{\obs} / 2}^{T_{\obs} / 2} dt \, e^{2\pi i f_j t} \Delta {T}_a(t) \,. 
    \label{est1}
\end{align}
We assume that $\Delta {T}_a(t)$ consists of contributions from the vector DM $\Delta {T}^{\text{VDM}}_a(t)$, stochastic gravitational waves $\Delta {T}^{\text{GW}}_a(t)$, and noise ${n}_a(t)$:
\begin{align}
    \Delta {T}_a(t) = \Delta {T}^{\text{VDM}}_a(t) + \Delta {T}^{\text{GW}}_a(t) + {n}_a(t) \,. 
    \label{est2}
\end{align}
For simplicity, we assume that the noise is uncorrelated between different pulsars, and the variance of the noise is common to all pulsars. We write $\braket{n_a(t) \, n_b(t)} \equiv \sigma^2 \delta_{ab}$, where the angle bracket denotes the ensemble average and $\delta_{ab}$ is the Kronecker delta. Assuming white noise for further simplification, we have $\braket{ \hat{n}_a(f_j) \, \hat{n}_b^*(f_k) } \approx \sigma^2 \delta_{ab} \delta_{jk} \cdot \Delta t / T_{\obs}$, where the asterisk (*) denotes the complex conjugate, and $\Delta t$ is the observation cadence.
We take $\sigma \sim 100 \, \text{nsec}$ and $\Delta t \sim \text{a few weeks}$ as typical values.

In the following, we focus on a specific frequency bin $f_A \equiv m/\pi$, to which the DM with a mass $m$ contributes.
The parameter estimation procedure can be constructed as follows (see, e.g., Sec.\,40 of Ref.~\cite{ParticleDataGroup:2024cfk}).
From the timing residuals of the pulsars $a$ and $b$, we consider the following cross product,
\begin{align}
    y_{ab} \equiv \Delta \hat{T}_a (f_A) \, \Delta \hat{T}^{*}_b (f_A) \,. 
    \label{est3}
\end{align}
Let $\bm{\theta} = (\theta_1, \theta_2, \dots)$ denote the set of parameters of the theory (or the model) to be estimated, and let $\mu_{ab}(\bm{\theta})$ be the theoretical template of $y_{ab}$ for given parameters $\bm{\theta}$. 
Then, the best-fit values of $\bm{\theta}$ can be obtained by minimizing the following $\chi^2$ function,
\begin{align}
    \chi^2 (\bm{\theta}) \equiv \sum_{ab, cd} (y_{ab} - \mu_{ab}(\bm{\theta})) \, (V^{-1})_{ab, cd} \, (y_{cd} - \mu_{cd}(\bm{\theta})) \,,
    \label{est4}
\end{align}
where the sum is taken over all pulsar pairs $(a,b)$ and $(c,d)$, and $(V^{-1})_{ab, cd}$ is the inverse of the covariance matrix $V_{ab, cd} \equiv \text{cov}[y_{ab},y_{cd}^*] \equiv \braket{y_{ab} \, y_{cd}^*} - \braket{y_{ab}} \braket{y_{cd}^*}$.
Furthermore, the statistical error of the parameters $\Delta \theta_i$ can be estimated from
\begin{align}
    \braket{ \Delta \theta_i \, \Delta \theta_j } = (\Gamma^{-1})_{ij} \,.
    \label{est5}
\end{align}
Here, $(\Gamma^{-1})_{ij}$ is the inverse of the Fisher matrix $\Gamma_{ij}$ given by 
\begin{align}
    \Gamma_{ij} \equiv \frac{1}{2} \frac{\pd^2 \chi^2(\bm{\theta})}{\pd \theta_i \, \pd \theta_j}\,, 
    \label{est6}
\end{align}
with the derivatives evaluated at the parameter values that minimize $\chi^2(\bm{\theta})$.

We define the template $\mu_{ab}(\bm{\theta})$ as the statistical mean of $y_{ab}$ for given parameters $\bm{\theta}$.
In our case \eqref{est2}, this consists of contributions from the vector DM, $\mu^{\text{VDM}}_{ab} = \braket{ \Delta \hat{T}^{\text{VDM}}_a (f_A) \, \Delta \hat{T}^{\text{VDM}*}_b (f_A) }$, and stochastic gravitational waves, $\mu^{\text{GW}}_{ab} = \braket{ \Delta \hat{T}^{\text{GW}}_a (f_A) \, \Delta \hat{T}^{\text{GW}*}_b (f_A) }$.
Using the results in Sec.~\ref{sec:angular_cor}, we obtain the vector DM contribution as\footnote{
While the angle bracket $\braket{\cdots}$ in Sec.~\ref{sec:angular_cor} includes the average over pulsar pairs for a given vector DM state with a fixed polarization plane, here it refers to the statistical average over vector DM realizations for a given pulsar pair. Specifically, here we consider the average over the polarization axis directions of the vector DM with a fixed amplitude $\bar{A}$ and polarization angle $\beta$. The two results are equivalent under the assumption of uniformly distributed pulsars.
}
\begin{align}
    \mu^{\text{VDM}}_{ab} = \frac{1}{2} \Phi_{\text{VDM}} \, \Gamma_{\text{VDM}}(\xi_{ab}; \beta) \,, 
    \label{est7}
\end{align}
where $\Phi_{\text{VDM}}$ and $\Gamma_{\text{VDM}}$ are defined in Eqs.~\eqref{Phivector} and \eqref{Gammavector}, respectively.
The $\xi_{ab}$ denotes the separation angle between the pulsars $a$ and $b$, and $\beta$ is the polarization angle of the vector DM at the Earth (denoted by $\beta(\bm{0})$ in the main text).
On the other hand, the correlation created by stochastic gravitational waves is expressed as 
\begin{align}
    \mu^{\text{GW}}_{ab} = \frac{1}{2} \Phi_{\text{GW}}(f_A) \, \Gamma_{\text{HD}}(\xi_{ab}) \,, 
    \label{est8}
\end{align}
where $\Phi_{\text{GW}}$ and $\Gamma_{\text{HD}}$ are given by Eqs.~\eqref{PhiGW1} and \eqref{HD}, respectively.
To evaluate $\chi^2(\bm{\theta})$, we also need the covariance matrix $V_{ab,cd}$. Assuming that the noise $n_a(t)$ is dominant in $\Delta T_a(t)$ and is Gaussian, we obtain 
\begin{align}
    V_{ab,cd} &= \braket{ \hat{n}_a(f_A) \, \hat{n}_b^*(f_A) \, \hat{n}_c^*(f_A) \, \hat{n}_d(f_A) } - 
    \braket{ \hat{n}_a(f_A) \, \hat{n}_b^*(f_A) } \braket{\hat{n}_c^*(f_A) \, \hat{n}_d(f_A) } 
    \notag \\
    &= \delta_{ac} \delta_{bd}\,  (\sigma^2 \cdot \Delta t / T_{\obs})^2 \,. 
    \label{est9}
\end{align}
Hence, the expression for $\chi^2 (\bm{\theta})$ in Eq.~\eqref{est4} becomes
\begin{align}
    \chi^2(\bm{\theta}) = \frac{1}{(\sigma^2 \cdot \Delta t / T_{\obs})^{2}} \sum_{ab} (y_{ab} - \mu_{ab}(\bm{\theta}))^2 \,,
    \label{est10}
\end{align}
where $\mu_{ab} = \mu_{ab}^{\text{VDM}} + \mu_{ab}^{\text{GW}}$.

\begin{figure}[tb]
\centering
\includegraphics[width=0.45\textwidth]{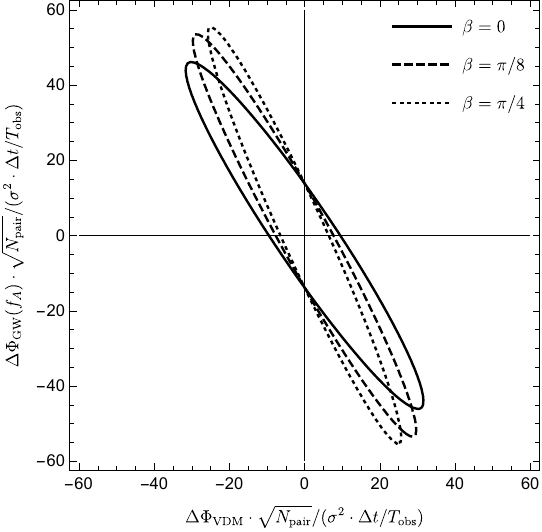}
\caption{The error contour for the two parameters $\Phi_{\text{VDM}}$ and $\Phi_{\text{GW}}(f_A)$ is shown. The solid, dashed, and dotted curves correspond to the cases of vector DM with $\beta = 0$ (linear polarization), $\beta = \pi/8$, and $\beta = \pi/4$ (perfect circular polarization), respectively.}
\label{fig:errorcontour}
\end{figure}

For simplicity, we treat $\Phi_{\text{VDM}} \equiv \theta_1$ and $\Phi_{\text{GW}}(f_A) \equiv \theta_2$ as the two parameters to be estimated, while keeping the polarization angle $\beta$ fixed.
Then, the components of the matrix $\Gamma_{ij}$ defined in Eq.~\eqref{est6} are given by 
\begin{align}
    \Gamma_{11} &= \frac{1}{(\sigma^2 \cdot \Delta t / T_{\obs})^{2}} \sum_{ab} \bigg[ \frac{1}{2} \Gamma_{\text{VDM}}(\xi_{ab}; \beta) \bigg]^2 
    \approx 
    \frac{N_{\text{pair}}}{(\sigma^2 \cdot \Delta t / T_{\obs})^{2}} \frac{1}{2} \int_{-1}^1 d \cos \xi \, \bigg[ \frac{1}{2} \Gamma_{\text{VDM}}(\xi; \beta) \bigg]^2 
    \label{est11}\\
    \Gamma_{12} = \Gamma_{21} &= \frac{1}{(\sigma^2 \cdot \Delta t / T_{\obs})^{2}} \sum_{ab} \frac{1}{2} \Gamma_{\text{VDM}}(\xi_{ab}; \beta) \cdot \frac{1}{2} \Gamma_{\text{HD}}(\xi_{ab}) 
    \approx 
     \frac{N_{\text{pair}}}{(\sigma^2 \cdot \Delta t / T_{\obs})^{2}} 
     \frac{1}{2} \int_{-1}^1 d\cos \xi \, \bigg[ \frac{1}{2} \Gamma_{\text{VDM}}(\xi; \beta) \cdot \frac{1}{2} \Gamma_{\text{HD}}(\xi) \bigg]
    \,,  
    \label{est12}\\
    \Gamma_{22} &= \frac{1}{(\sigma^2 \cdot \Delta t / T_{\obs})^{2}} \sum_{ab} \bigg[ \frac{1}{2} \Gamma_{\text{HD}}(\xi_{ab}) \bigg]^2 
    \approx \frac{N_{\text{pair}}}{(\sigma^2 \cdot \Delta t / T_{\obs})^{2}} \frac{1}{2} \int_{-1}^1 d\cos \xi \, \bigg[ \frac{1}{2} \Gamma_{\text{HD}}(\xi) \bigg]^2 
    \,,
    \label{est13}
\end{align}
 where, assuming uniformly distributed pulsars, the sum over pulsar pairs is replaced by an integral over the separation angle $\xi$, and $N_{\text{pair}}$ denotes the total number of pulsar pairs.
The integrals can be evaluated analytically using the explicit forms of $\Gamma_{\text{VDM}}$ and $\Gamma_{\text{HD}}$ given in Eqs.~\eqref{Gammavector} and \eqref{HD}, yielding
 \begin{align}
     \begin{pmatrix}
         \Gamma_{11} & \Gamma_{12} \\
         \Gamma_{21} & \Gamma_{22}
     \end{pmatrix}
     =
     \frac{N_{\text{pair}}}{(\sigma^2 \cdot \Delta t / T_{\obs})^{2}} 
     \begin{pmatrix}
         \tilde{\Gamma}_{11} & \tilde{\Gamma}_{12} \\
         \tilde{\Gamma}_{21} & \tilde{\Gamma}_{22}
     \end{pmatrix}
     \label{est14}
 \end{align}
 with 
 \begin{align}
     \tilde{\Gamma}_{11} 
     &= \frac{17533}{1015680} - \frac{111}{16928} \cos 4 \beta + \frac{383}{1015680} \cos 8\beta \,, 
     \label{est15} \\
     \tilde{\Gamma}_{12} = \tilde{\Gamma}_{21}
     &= \frac{5}{552} - \frac{1}{552} \cos 4\beta \,, 
     \label{est16} \\
     \tilde{\Gamma}_{22} &= \frac{1}{192} \,.
     \label{est17}
 \end{align}
The (co)variances of the parameters $\Phi_{\text{VDM}}$ and $\Phi_{\text{GW}}(f_A)$ are then estimated from Eq.~\eqref{est5} as
\begin{align}
    \begin{pmatrix}
        \braket{ (\Delta \Phi_{\text{VDM}})^2 } &\braket{ \Delta \Phi_{\text{VDM}} \, \Delta \Phi_{\text{GW}}(f_A) }     \\
        \braket{ \Delta \Phi_{\text{GW}}(f_A) \, \Delta \Phi_{\text{VDM}} } 
        &\braket{ (\Delta \Phi_{\text{GW}}(f_A))^2 }
    \end{pmatrix}
    &= 
    \begin{pmatrix}
        (\Gamma^{-1})_{11} &(\Gamma^{-1})_{12} \\
        (\Gamma^{-1})_{21} &(\Gamma^{-1})_{22} \\
    \end{pmatrix}
    = \frac{(\sigma^2 \cdot \Delta t / T_{\obs})^{2}}{N_{\text{pair}}}     \begin{pmatrix}
        (\tilde{\Gamma}^{-1})_{11} &(\tilde{\Gamma}^{-1})_{12} \\
        (\tilde{\Gamma}^{-1})_{21} &(\tilde{\Gamma}^{-1})_{22} \\
    \end{pmatrix}\,,
    \label{est18}
\end{align}
where $(\tilde{\Gamma}^{-1})_{ij}$ is the inverse of the matrix $\tilde{\Gamma}_{ij}$, which depends on the polarization angle $\beta$.
Figure \ref{fig:errorcontour} shows the error contours specified by $\Gamma_{ij} \, \Delta \theta_i \, \Delta \theta_j = 1$ for several values of $\beta$.
It is evident that the error in the $\Phi_{\text{VDM}}$-direction is minimized when $\beta = \pi/4$, corresponding to the perfectly circularly polarized vector DM.
This is because the pulsar correlation is maximally enhanced at $\beta=\pi/4$ for a fixed DM amplitude (see Fig.~\ref{fig:angular}).
However, Fig.~\ref{fig:errorcontour} also shows that the error in $\Phi_{\text{GW}}(f_A)$ simultaneously increases.
This can be understood from the fact that the quadrupole component of the vector DM correlation becomes more pronounced and increasingly degenerate with the gravitational wave contribution as $\beta \to \pi/4$.

In terms of the total number of pulsars $N_{\text{pul}}$, the number of independent pulsar pairs is given by $N_{\text{pair}} = N_{\text{pul}}(N_{\text{pul}} - 1)/2 \sim N_{\text{pul}}^2/2$ for $N_{\text{pul}} \gg 1$.
Then, the standard deviation (i.e., the square root of the variance) of $\Phi_{\text{VDM}}$ is estimated as 
\begin{align}
    \sqrt{\braket{ (\Delta \Phi_{\text{VDM}})^2 }}
    \sim 1 \times 10^{-17}\, \text{sec}^2 \bigg( \frac{10^2}{N_{\text{pul}}} \bigg) 
    \bigg( \frac{\sigma}{100\,\text{nsec}} \bigg)^2 
    \bigg( \frac{\Delta t}{\text{yr}/20} \bigg) 
    \bigg( \frac{15\,\text{yr}}{T_{\obs}} \bigg) 
    \bigg( \frac{(\tilde{\Gamma}^{-1})_{11}}{10^3} \bigg)^{1/2}\,, 
    \label{est19}
\end{align}
which should be compared with the mean value of $\Phi_{\text{VDM}}$ given by Eq.~\eqref{Phivector2}.
We see that the standard deviation $\sqrt{\braket{ (\Delta \Phi_{\text{VDM}})^2 }}$ remains smaller than $\Phi_{\text{VDM}}$ for $m \lesssim 10^{-23}\, \text{eV}$, assuming fiducial parameter values for the experiment, although the signal-to-noise ratio must be sufficiently large for the detection when we take into account the look-elsewhere effects. In principle, the estimation error on 
$\Phi_{\text{VDM}}$ can be further reduced even for larger DM masses by increasing the number of pulsars $N_{\text{pul}}$ and lowering the noise level $\sigma$.

\bibliography{refs_VectorDM_PTA}% Produces the bibliography via BibTeX.

\end{document}